\documentclass[useAMS,referee]{biom}

\usepackage{moreverb}
\usepackage{bm}
\usepackage{graphicx,psfrag,epsf, float}
\usepackage{setspace}
\usepackage{color}
\newcommand{\rev}[1]{{\color{black}#1}}

\usepackage{tikz}
\usepackage{algorithm}
\usepackage{algpseudocode}

\usetikzlibrary{shapes,decorations,arrows,calc,arrows.meta,fit,positioning}
\tikzset{
    -Latex,auto,node distance =1 cm and 1 cm,semithick,
    state/.style ={ellipse, draw, minimum width = 0.7 cm},
    point/.style = {circle, draw, inner sep=0.04cm,fill,node contents={}},
    bidirected/.style={Latex-Latex,dashed},
    el/.style = {inner sep=2pt, align=left, sloped}
}

\usepackage{amsfonts}
\usepackage{hyperref}
\hypersetup{colorlinks=true, citecolor=blue, linkcolor=blue, filecolor=blue}
\usepackage{url}
\usepackage{mathtools}
\usepackage{enumerate, color}
\usepackage{amsmath}
\usepackage{algorithm, algorithmicx, algpseudocode}
\usepackage{graphicx}
\usepackage{lscape}
\usepackage{natbib}
\usepackage{xr}

\usepackage{enumitem}
\setlist{itemsep=1pt, topsep=2pt, parsep=0pt, partopsep=0pt}
\usepackage{enumitem}

\def\bSig\boldsymbol{\Sigma}

\title[Selective Borrowing under Covariate Mismatch]{Selective Information Borrowing for Region-Specific Treatment Effect Inference under Covariate Mismatch in Multi-Regional Clinical Trials}

\author{Chenxi Li$^{1}$, 
Ke Zhu$^{1,2}$, Shu Yang$^{2}$ and Xiaofei Wang$^{1,*}$\email{xiaofei.wang@duke.edu} \\
$^{1}$Department of Biostatistics and Bioinformatics, Duke University, Durham, NC 27710, U.S.A. \\
$^{2}$Department of Statistics, North Carolina State University, Raleigh, NC 27695, U.S.A.}

\setlist{nosep}
\setlist[itemize]{topsep=2pt,partopsep=0pt,itemsep=1pt,parsep=0pt,leftmargin=*}
\setlist[enumerate]{topsep=2pt,partopsep=0pt,itemsep=1pt,parsep=0pt,leftmargin=*}

\begin{document}

%  This will produce the submission and review information that appears
%  right after the reference section.  Of course, it will be unknown when
%  you submit your paper, so you can either leave this out or put in 
%  sample dates (these will have no effect on the fate of your paper in the
%  review process!)

%\date{{\it Received XX} 20XX. {\it Revised XX} 20XX.  {\it Accepted XX} 20XX.}

%  These options will count the number of pages and provide volume
%  and date information in the upper left hand corner of the top of the 
%  first page as in published papers.  The \pagerange command will only
%  work if you place the command \label{firstpage} near the beginning
%  of the document and \label{lastpage} at the end of the document, as we
%  have done in this template.

%  Again, putting a volume number and date is for your own amusement and
%  has no bearing on what actually happens to your paper!  

\pagerange{\pageref{firstpage}--\pageref{lastpage}} 
\volume{}
\pubyear{}
\artmonth{}

%  The \doi command is where the DOI for your paper would be placed should it
%  be published.  Again, if you make one up and stick it here, it means 
%  nothing!

\doi{}

%  This label and the label ``lastpage'' are used by the \pagerange
%  command above to give the page range for the article.  You may have 
%  to process the document twice to get this to match up with what you 
%  expect.  When using the referee option, this will not count the pages
%  with tables and figures.  

\label{firstpage}

%  put the summary for your paper here

\begin{abstract}
Multi-regional clinical trials (MRCTs) are central to global drug development, enabling evaluation of treatment effects across diverse populations. A key challenge is valid and efficient inference for a region-specific estimand when the target region is small and differs from auxiliary regions in baseline covariates or unmeasured factors. We adopt an estimand-based framework and focus on the region-specific average treatment effect (RSATE) in a prespecified target region, which is directly relevant to local regulatory decision-making. Cross-region differences can induce covariate shift, covariate mismatch, and outcome drift, potentially biasing information borrowing and invalidating RSATE inference. To address these issues, we develop a unified causal inference framework with selective information borrowing. First, we introduce an inverse-variance weighting estimator that combines a “small-sample, rich-covariate” target-only estimator with a “large-sample, limited-covariate” full-borrowing doubly robust estimator, maximizing efficiency under no outcome drift. Second, to accommodate outcome drift, we apply conformal prediction to assess patient-level comparability and adaptively select auxiliary-region patients for borrowing. Third, to ensure rigorous finite-sample inference, we employ a conditional randomization test with exact, model-free, selection-aware type I error control. Simulation studies show the proposed estimator improves efficiency, yielding 10–50\% reductions in mean squared error and higher power relative to no-borrowing and full-borrowing approaches, while maintaining valid inference across diverse scenarios. An application to the POWER trial further demonstrates improved precision for RSATE estimation.

%Together, these findings demonstrate that our framework offers a robust and principled solution for RSATE evaluation in MRCTs and can directly support local regulatory decision-making.

\vspace{5pt}
\end{abstract}

%  Please place your key words in alphabetical order, separated
%  by semicolons, with the first letter of the first word capitalized,
%  and a period at the end of the list.
%


%  As usual, the \maketitle command creates the title and author/affiliations
%  display 

\maketitle

%  If you are using the referee option, a new page, numbered page 1, will
%  start after the summary and keywords.  The page numbers thus count the
%  number of pages of your manuscript in the preferred submission style.
%  Remember, ``Normally, regular papers exceeding 25 pages and Reader Reaction 
%  papers exceeding 12 pages in (the preferred style) will be returned to 
%  the authors without review. The page limit includes acknowledgements, 
%  references, and appendices, but not tables and figures. The page count does 
%  not include the title page and abstract. A maximum of six (6) tables or 
%  figures combined is often required.''

%  You may now place the substance of your manuscript here.  Please use
%  the \section, \subsection, etc commands as described in the user guide.
%  Please use \label and \ref commands to cross-reference sections, equations,
%  tables, figures, etc.
%
%  Please DO NOT attempt to reformat the style of equation numbering!
%  For that matter, please do not attempt to redefine anything!

%\doublespacing
\vspace{-30pt}
\section{Introduction} \label{introduction}

Multi-regional clinical trials (MRCTs) have emerged as a vital strategy in global drug development, enabling sponsors to evaluate treatment efficacy and safety across diverse  populations. By enrolling patients from multiple geographically separated regions, MRCTs improve trial efficiency, accelerate regulatory submissions, and facilitate faster access to new therapies worldwide \citep{chen2009bayesian,Chen2010,bean2021,bean2023bayesian,zhuang2024,hua2024inference,bean2024bayesian,Robertson2025,dette2025testing,wan2025estimation,alene2025analyzing}. In addition to estimating the global effect of an active treatment, another main objective of MRCTs is to quantify region-specific treatment effects to support local registration \citep{ICH1998,Hung2010}. This is especially important for regulatory review, where local health authorities increasingly emphasize the need for region-specific evidence to justify approval decisions when treatment effects vary across participating countries \citep{ICH_E17, Guo2016}. A key challenge arises when patient recruitment is difficult in the target region or when regional subpopulations are small. With limited sample sizes, estimating treatment effects using only region-specific data can be inefficient, motivating the need to borrow information from other regions to improve inference for the target population. From a formal estimand perspective, these questions are naturally framed in terms of a region-specific average treatment effect in the target-region population, which explicitly links the trial design and analysis to the local regulatory decision problem.

However, when borrowing information across regions, it is critical to account for regional heterogeneity to avoid biased borrowing and invalid inference. \cite{ICH1998} acknowledges that data from one region may not be directly extrapolated to another because of intrinsic differences such as ethnicity or genetic background, and \cite{ICH_E17} further recommends prespecifying borrowing or shrinkage strategies and using covariate adjustment for baseline prognostic factors. In practice, these considerations raise three major challenges. First, baseline covariate shift and mismatch are common. Regions often have substantially different covariate distributions and, in some cases, nonoverlapping covariate sets, since the availability or definition of baseline variables can differ due to healthcare infrastructure, data collection protocols, or ethical constraints. Naively borrowing information with such discrepancies induces bias, while restricting analyses to shared covariates may reduce efficiency, especially when additional region-specific covariates in the target region have high predictive power. Second, even with balanced measured covariates, unmeasured factors such as clinical practice patterns may still vary across regions, creating hidden bias between the target and auxiliary regions. This issue is often referred to as outcome drift. Third, valid inference remains difficult. Limited target-region sample sizes can undermine large-sample approximations; covariate-shift adjustments relying on propensity score or outcome models may fail when both models are misspecified; and dynamic borrowing procedures designed for outcome drift can yield invalid inference if selection uncertainty is not properly incorporated.

To address the challenges that arise during information borrowing, a broad set of methods has been developed in the data integration literature \citep{colnet2024causal}. For covariate shift between data sources, propensity score matching and weighting, outcome regression, and doubly robust estimators have been proposed \citep{Valancius2024}. The covariate mismatch problem has received growing attention \citep{han2023multiply,han2024improving,zeng2025efficient,li2025generalizing,williams2025nonparametric}, and is also referred to as block missing in the transfer learning literature \citep{chang2024heterogeneous,xu2025representation}. For outcome drift, Bayesian dynamic borrowing approaches adaptively downweight source information according to its similarity to the target data, including informative priors \citep{chen2000power,hobbs2011hierarchical,schmidli2014robust,yang2023sam,kwiatkowski2024,Alt2024} and Bayesian hierarchical models \citep{thall2003hierarchical,kaizer2018bayesian}. Recently, several frequentist strategies robust to outcome drift have been proposed, including prognostic adjustment or digital twins \citep{schuler2022increasing,liu2025coadvise}, test-then-pool \citep{viele2014use,yang2023elastic,gao2023pretest,dang2022cross}, selective borrowing \citep{gao2025improving,gao2024,Zhu2024,liu2025robust}, bias correction \citep{stuart2008matching,wu2022integrative,yang2024datafusion,mao2025statistical,ye2025integrative,van2024adaptive}, aggregate combination \citep{chen2021minimax,cheng2021adaptive,rosenman2023combining}, and approaches that leverage auxiliary variables \citep{guo2022multisource,dang2022cross}. For valid inference after borrowing, Bayesian procedures typically require additional calibration to control type I error, whereas frequentist approaches include several options: (i) asymptotic normality assuming no outcome drift \citep{Li2023,Valancius2024} or under oracle selection of unbiased sources \citep{gao2025improving} or under oracle correction of bias \citep{yang2024datafusion,van2024adaptive}, (ii) an asymptotic non-normal distribution that account for selection uncertainty \citep{yang2023elastic,dang2022cross}, or (iii) randomization or permutation tests that incorporate selection uncertainty \citep{Zhu2024,liu2025robust,ren2025leveraging}. In the MRCT setting, \cite{Robertson2025} proposed a target-only estimator without borrowing and a doubly robust estimator that borrows all information when no outcome drift is present, and then selected between them using a pretest of region-outcome associations, essentially following a test-then-pool strategy; their inference relies on asymptotic normality without accounting for the uncertainty introduced by the pretest.

Building on these developments, and focusing on the region-specific average treatment effect (RSATE) as the formal estimand for the target region, we make three contributions in this paper. First, we propose an inverse variance weighting (IVW) method that combines the target-only estimator (small sample, rich covariates) with the full-borrowing doubly robust estimator (large sample, limited covariates) to maximize efficiency under the assumption of no outcome drift. Second, recognizing that the outcome drift assumption may be violated when using the auxiliary region data as a whole, we assess the individual comparability of each patient using conformal prediction \citep{vovk2005algorithmic,lei2018distribution,angelopoulos2023conformal}, adaptively determine a threshold to select a subset of patients who satisfy the assumption of no outcome drift, and then apply the IVW estimator to the target region data together with this selected subset of auxiliary region data. Third, to facilitate valid inference, we employ the conditional randomization test to deliver finite-sample exact, model-free, and selection-aware inference \citep{fisher1935,Zheng2008,Zhang2023}. Taken together, these components yield a robust and efficient estimand-based framework for inferring region-specific average treatment effects with selective information borrowing that addresses covariate shift and mismatch, outcome drift, and finite-sample type I error control, and is directly aligned with local regulatory assessment in MRCTs.

The paper proceeds as follows. Section \ref{Framework} introduces the general causal inference framework for MRCTs. Section \ref{sec:mismatch} describes information borrowing methods under covariate shift and mismatch. Section \ref{sec:CSB} presents the proposed CSB method. Section \ref{FRT} presents a randomization inference framework to control the type I error rate while strictly enhancing power. Section \ref{Simulation} shows the design and results of simulation studies. Section \ref{Realdata} applies the proposed method to POWER trial data. We conclude with a discussion in Section \ref{Discussion}.

\vspace{-30pt}
\section{Preliminary}\label{Framework}
\vspace{-10pt}
\subsection{Target region, estimand, and data structure}

%target region R
%potential outcome Y(a), estimand tau
%design A, observed Y
%covarite X and U

We use the POWER trial as a motivating example to introduce notation and the potential outcome framework. The trial enrolls patients from multiple geographically distinct regions, including North America, South America, and Europe, with a total sample size of $n$. Rather than targeting the pooled population across all regions, our primary interest lies in a prespecified target region, for example North America, which aligns with the objective of supporting local regulatory decision making.
Accordingly, we partition the study sample into $n_{\mathcal{R}}$ patients from the \textit{target region} with $R=1$ and $n_{\mathcal{E}}$ patients from \textit{auxiliary regions} with $R=0$, where $R$ denotes the sample indicator. The auxiliary regions include all non-target regions, such as South America and Europe. We do not further distinguish among auxiliary regions, since heterogeneity will be handled at the individual-patient level in Section~\ref{sec:CSB}. 

The treatment comparison of interest is between the active treatment $A=1$ (enobosarm) and the control $A=0$ (placebo). Let $Y(a)$ denote the potential outcome under treatment $a$, defined as the percentage change in lean body mass at Day 84. The estimand of interest is the region-specific average treatment effect (RSATE), defined as $\tau=\theta_1-\theta_0$, where $\theta_a=\mathbb{E}[Y(a)\mid R=1]$ denotes the mean potential outcome under treatment $a$ in the target-region population. In the POWER trial, the $n$ patients are randomized to the treatment group $A=1$ with sample size $n_1$ or to the control group $A=0$ with sample size $n_0$, where $A$ is the treatment indicator and the observed outcome is denoted by $Y$.

\begin{figure}[htbp]
    \centering   \includegraphics[width=\textwidth]
    {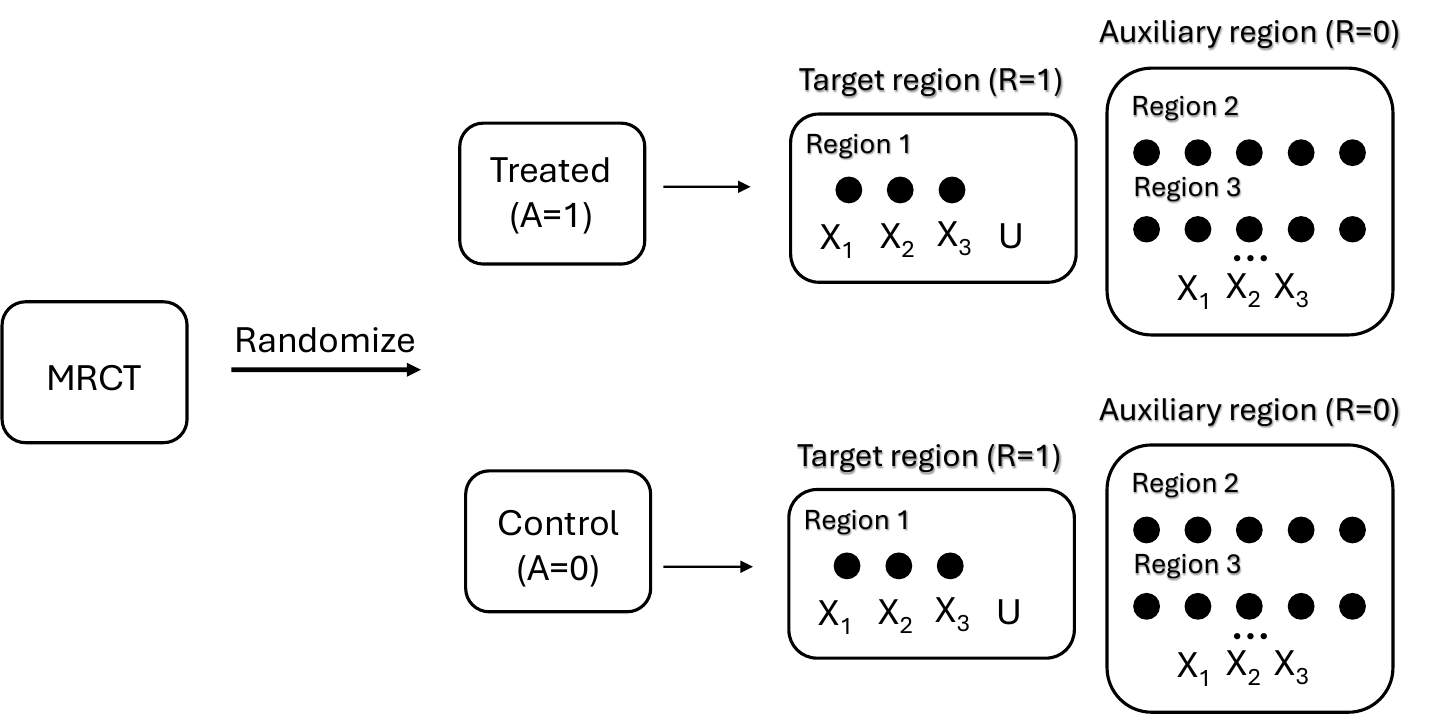}
    \caption{Data structure of a multi-regional randomized clinical trial with target and auxiliary regions.}
    \label{data_structure}
\end{figure}

Across all regions, a set of common baseline covariates, such as age, sex, and ECOG performance status, is collected and denoted by $X$. 
In general, the target region may share different sets of covariates with different auxiliary regions; we consider this case in Appendix F of the Supplementary Materials.
In addition, the target region may include region-specific baseline covariates $U$ that are unavailable or not directly comparable in auxiliary regions due to differences in data collection, definitions, or measurement practices. For example, a weight-loss flag is recorded in the North American region but not in auxiliary regions. The observed data structure is therefore $\{R_i, X_i, U_i, A_i, Y_i\}_{i=1}^n$, where $U_i$ is missing for individuals with $R_i=0$. Table~\ref{observed} and Figure~\ref{data_structure} summarizes the observed data structure.
 
\vspace{-10pt}
\begin{table}[t]
    \caption{Observed data structure. “\checkmark
” and “?” indicate observed and unobserved, respectively}
    \centering
    \begin{tabular}{ccccccccc}
        %\toprule
        \Hline
        & \textbf{ID} & \textbf{Treatment} & \multicolumn{2}{c}{\textbf{Covariates}} & \multicolumn{2}{c}{\textbf{Potential Outcome}} & \textbf{Observed Outcome} \\ 
        & \  & $A$ & ${X}$ & ${U}$ & $Y(1)$ & $Y(0)$ & $Y$ \\ \hline
        %\midrule
        Target & 1 & 1 & \checkmark & \checkmark & \checkmark & ? & \checkmark \\
        & $\vdots$ & $\vdots$ & $\vdots$ & $\vdots$ & $\vdots$ & $\vdots$ & $\vdots$ \\
        & $n_\mathcal{R}$ & 0 & \checkmark & \checkmark & ? & \checkmark & \checkmark \\ \hline
        %\midrule
        Auxiliary & $n_{\mathcal{R}}+1$ & 1 & \checkmark & ? & \checkmark & ? & \checkmark \\
        & $\vdots$ & $\vdots$ & $\vdots$ & $\vdots$ & $\vdots$ & $\vdots$ & $\vdots$ \\
        & $n$ & 0 & \checkmark & ? & ? & \checkmark & \checkmark \\
        \hline
       % \bottomrule
    \end{tabular}
    \label{observed}
\end{table}
\vspace{-20pt}

\vspace{-10pt}
\subsection{Target-region-only identification and estimation}

%identification assumption
%NB-Xonly
%NB-AllCov

We first introduce the following identification assumptions  \citep{Robertson2025}.
\vspace{-15pt}
\begin{assumption}
\label{ass:rct}
(i) (Consistency) $Y = A Y(1) + (1-A)Y(0)$. 
(ii) (Unconfoundedness) $Y(a) \perp A \mid X, R=1$ for $a\in\{0,1\}$. 
(iii) (Positivity) $0<\mathbb{P}(A=1\mid X, R=1)<1$.
\end{assumption}
\vspace{-10pt}

Assumption \ref{ass:rct} is typically satisfied by the design of MRCTs. Part (i) requires no interference between individuals and no multiple versions of treatment that would lead to different outcomes for the same individual. Part (ii) assumes the absence of unmeasured confounding within the target region after conditioning on baseline covariates. In MRCTs, treatment assignment is usually independent of covariates and regions by design, or independent of covariates within regions when region-stratified randomization is employed. Here, we impose a weaker condition that only requires independence between potential outcomes and treatment assignment given baseline covariates within the target region. Part (iii) requires that each individual in the target region has a positive probability of receiving either treatment level given covariates, which typically holds in MRCTs.

Under Assumption~\ref{ass:rct}, the RSATE $\tau$ is identifiable using data from the target region alone. Let $e_{a\mid1}(x)=\mathbb{P}(A=a\mid R=1,X=x)$ denote the treatment propensity score within the target region, which is known by design in MRCTs. Let $\mu_a(x)=\mathbb{E}[Y(a)\mid X=x,R=1]$ for $a\in\{0,1\}$ denote the conditional mean outcome functions based on the common baseline covariates $X$ in the target region. A covariate-adjusted estimator of $\tau$ that relies solely on target-region data and adjusts for $X$ is given by $\hat{\tau}^{\text{NB-Xonly}}=\hat{\theta}_1^{\text{NB-Xonly}}-\hat{\theta}_0^{\text{NB-Xonly}}$, where
$$
\hat{\theta}_a^{\text{NB-Xonly}}
=\frac{1}{n_{\mathcal{R}}}\sum_{i=1}^n \left[
R_i\hat{\mu}_a^{\text{NB}}(X_i)
+\frac{\mathbb{I}(R_i=1,A_i=a)}{e_{a\mid1}(X_i)}
\{Y_i-\hat{\mu}_a^{\text{NB}}(X_i)\}
\right].
$$
Here, $\hat{\mu}_a^{\text{NB}}(x)$ is a fitted outcome regression model for $\mu_a(x)$ using only observations from the target region with $R=1$. The superscript ``NB'' indicates ``no borrowing,'' as no information from auxiliary regions is incorporated.

The estimator $\hat{\tau}^{\text{NB-Xonly}}$ does not exploit target-region-specific covariates $U$, which are often strongly prognostic and can yield efficiency gains when adjusted for. Let $\mu_a(x,u)=\mathbb{E}[Y(a)\mid X=x,U=u,R=1]$ for $a\in\{0,1\}$ denote the conditional mean outcome functions based on all prognostic variables in the target region. 
%In practice, when additional target-region-specific covariates $U$ are available, selecting strongly prognostic variables can further improve efficiency and stability. This may be achieved using subject-matter knowledge or data-adaptive procedures such as penalized regression (e.g., adaptive lasso \citep{zou2006}) applied within the target region. 
We slightly abuse notation by using the same symbol $\mu_a$ and distinguishing the functions by their arguments. A covariate-adjusted estimator that incorporates both $X$ and $U$ is given by $\hat{\tau}^{\text{NB-AllCov}}=\hat{\theta}_1^{\text{NB-AllCov}}-\hat{\theta}_0^{\text{NB-AllCov}}$, where
$$
\hat{\theta}_a^{\text{NB-AllCov}}
=\frac{1}{n_{\mathcal{R}}}\sum_{i=1}^n \left[
R_i\hat{\mu}_a^{\text{NB}}(X_i,U_i)
+\frac{\mathbb{I}(R_i=1,A_i=a)}{e_{a\mid1}(X_i)}
\{Y_i-\hat{\mu}_a^{\text{NB}}(X_i,U_i)\}
\right].
$$
Here, $\hat{\mu}_a^{\text{NB}}(x,u)$ is a fitted outcome model for $\mu_a(x,u)$ using only target-region data, which may use data-adaptive procedures such as the adaptive lasso \citep{zou2006} or subject-matter knowledge to select strongly prognostic variables among $(X,U)$.

Both $\hat{\tau}^{\text{NB-Xonly}}$ and $\hat{\tau}^{\text{NB-AllCov}}$ exploit the treatment propensity score and are therefore robust to  outcome model misspecification. The estimator $\hat{\tau}^{\text{NB-AllCov}}$ is typically more efficient, as it adjusts for all prognostic variables in the target region. Under standard regularity conditions, both estimators are consistent and asymptotically normal, with variance estimators available in closed form; further details are provided in Appendix A.1. However, neither estimator borrows information from auxiliary regions, which motivates the borrowing strategies introduced in the next section to further improve efficiency.

\vspace{-10pt}
\subsection{Auxiliary-region-integrated identification and estimation}

%identification assumption
%FB-Xonly

To enable information borrowing from auxiliary regions, we introduce the following assumption \citep{Robertson2025}, which will be relaxed in Section~\ref{sec:CSB}.

\vspace{-20pt}

\begin{assumption}[Conditional mean exchangeability]
\label{ass:cme}
$\mathbb{E}[Y(a)\mid R=0,X]=\mathbb{E}[Y(a)\mid R=1,X]$ for $a\in\{0,1\}$.
\end{assumption}
\vspace{-10pt}
Assumption~\ref{ass:cme} allows outcome distributions to differ across regions but requires that any systematic difference between the target and auxiliary regions be fully explained by the commonly observed covariates $X$. In other words, after conditioning on $X$, the conditional mean potential outcomes are assumed to be invariant across regions. This assumption may fail in the presence of unmeasured covariates, such as $U$, that jointly affect regional membership and the outcome. We treat Assumption~\ref{ass:cme} as an idealized condition and develop information borrowing procedures under this assumption. In Section~\ref{sec:CSB}, we relax this requirement by allowing only a subset of auxiliary-region observations to satisfy conditional mean exchangeability, and we selectively borrow information from those observations.

Under Assumption~\ref{ass:cme}, the RSATE $\tau$ is identifiable using data from both the target and auxiliary regions. Let $e_a(x)=\mathbb{P}(A=a\mid X=x)$ denote the treatment propensity score across all regions, which is known by design in MRCTs. Let $\pi(x)=\mathbb{P}(R=1\mid X=x)$ denote the sampling propensity score, which is used to adjust for covariate shift between the target and auxiliary regions.
This quantity is typically unknown and is estimated by fitting a sampling model $R\sim X$ using data from all regions, with the fitted value denoted by $\hat{\pi}^{\text{FB}}(x)$. The superscript ``FB'' indicates ``full borrowing,'' as all auxiliary-region observations are incorporated under Assumption~\ref{ass:cme}.

A doubly robust estimator of $\tau$ that borrows information from all regions and adjusts for the covariates shift from $X$ is given by $\hat{\tau}^{\text{FB-Xonly}}=\hat{\theta}_1^{\text{FB-Xonly}}-\hat{\theta}_0^{\text{FB-Xonly}}$, where
$$
\hat{\theta}_a^{\text{FB-Xonly}}
=\frac{1}{n_{\mathcal{R}}}\sum_{i=1}^n\left[
R_i\,\hat{\mu}^{\text{FB}}_a(X_i)
+\hat{\pi}^{\text{FB}}(X_i)\frac{\mathbb{I}(A_i=a)}{e_a(X_i)}
\{Y_i-\hat{\mu}^{\text{FB}}_a(X_i)\}
\right].
$$
Here, $\hat{\mu}_a^{\text{FB}}(x)$ is a fitted outcome regression model for $\mu_a(x)$ constructed using observations from both the target and auxiliary regions.

Under Assumption~\ref{ass:cme}, the estimator $\hat{\theta}_a^{\text{FB-Xonly}}$ is doubly robust as it is consistent if either the outcome regression model or the sampling propensity score model is correctly specified, and it is semiparametrically efficient when both models are correctly specified. Under regularity conditions, $\hat{\theta}_a^{\text{FB-Xonly}}$ is asymptotically normal, with variance estimators available in closed form; further details are provided in Appendix A.2 in the Supplementary Materials.

\cite{Robertson2025} recommend using $\hat{\tau}^{\text{FB-Xonly}}$ when Assumption~\ref{ass:cme} holds, and using $\hat{\tau}^{\text{NB-AllCov}}$ when Assumption~\ref{ass:cme} fails. Their analysis does not consider the covariate mismatch problem, and no additional target-region-specific covariates $U$ are available in their setting. As a result, $\hat{\tau}^{\text{NB-AllCov}}$ reduces to $\hat{\tau}^{\text{NB-Xonly}}$ in their framework. While these recommendations are sensible under their assumptions, both can be further improved in settings where additional covariate information is available and exchangeability may only hold partially:
\vspace{-10pt}

\begin{itemize}[leftmargin=*]
    \item When Assumption~\ref{ass:cme} holds, the presence of additional covariates $U$ in the target region implies that $\hat{\theta}_a^{\text{FB-Xonly}}$ may be less efficient than $\hat{\theta}_a^{\text{NB-AllCov}}$, as the former ignores the predictive information contained in $U$. We address this limitation in Section~\ref{sec:mismatch} by developing estimators that incorporate $U$ while still borrowing information from auxiliary regions.
    \item When Assumption~\ref{ass:cme} fails globally but holds for a subpopulation of auxiliary region, it is suboptimal to discard all auxiliary-region information and rely solely on $\hat{\theta}_a^{\text{NB-AllCov}}$. Instead, we selectively borrow information from auxiliary observations that satisfy the exchangeability condition in Section~\ref{sec:CSB}.
\end{itemize}

\vspace{-30pt}
\section{Information borrowing under covariate mismatch}\label{sec:mismatch}

Under Assumption~\ref{ass:cme}, we further improve the estimator $\hat{\theta}_a^{\text{FB-Xonly}}$ by leveraging the additional predictive power of the target-region-specific covariates $U$. For target-region patients with $R_i=1$, there are two options for predicting its outcome $Y_i$: (i) $\hat{\mu}_a^{\text{NB}}(X_i,U_i)$, which exploits richer covariate information but is estimated using fewer observations, and (ii) $\hat{\mu}_a^{\text{FB}}(X_i)$, which uses a more limited set of covariates but is estimated from a larger pooled sample. Our main idea is to combine these two predictions using inverse variance weighting (IVW). 

Let the empirical prediction errors of $\hat{\mu}_a^{\text{NB}}(X_i,U_i)$ and $\hat{\mu}_a^{\text{FB}}(X_i)$ be given by
$$
v_{\text{NB}}
=\frac{\sum_{i=1}^n \mathbb{I}(A_i=a,R_i=1)\{Y_i-\hat{\mu}_a^{\text{NB}}(X_i,U_i)\}^2}
{\sum_{i=1}^n \mathbb{I}(A_i=a,R_i=1)},
\quad
v_{\text{FB}}
=\frac{\sum_{i=1}^n \mathbb{I}(A_i=a)\{Y_i-\hat{\mu}_a^{\text{FB}}(X_i)\}^2}
{\sum_{i=1}^n \mathbb{I}(A_i=a)}.
$$
The IVW prediction is then defined as
$$
\hat{Y}_i^{\text{FB}}
=
\begin{cases}
\dfrac{v_{\text{FB}}}{v_{\text{NB}}+v_{\text{FB}}}\hat{\mu}_a^{\text{NB}}(X_i,U_i)
+\dfrac{v_{\text{NB}}}{v_{\text{NB}}+v_{\text{FB}}}\hat{\mu}_a^{\text{FB}}(X_i),
& R_i=1, \\
\hat{\mu}_a^{\text{FB}}(X_i),
& R_i=0.
\end{cases}
$$
Since the target-region-specific covariates $U$ are unavailable for auxiliary-region observations, the prediction for individuals with $R_i=0$ relies solely on $\hat{\mu}_a^{\text{FB}}(X_i)$.

Replacing $\hat{\mu}_a^{\text{FB}}(X_i)$ with $\hat{Y}_i^{\text{FB}}$ in $\hat{\theta}_a^{\text{FB-Xonly}}$ yields $\hat{\tau}^{\text{FB-IVW}}=\hat{\theta}_1^{\text{FB-IVW}}-\hat{\theta}_0^{\text{FB-IVW}}$, where
\begin{equation}
\label{eq:FB-IVW}
\hat{\theta}_a^{\text{FB-IVW}}
=\frac{1}{n_{\mathcal{R}}}\sum_{i=1}^n\left[
R_i\,\hat{Y}_i^{\text{FB}}
+\hat{\pi}^{\text{FB}}(X_i)\frac{\mathbb{I}(A_i=a)}{e_a(X_i)}
\{Y_i-\hat{Y}_i^{\text{FB}}\}
\right].
\end{equation}
Under Assumptions~\ref{ass:rct} and \ref{ass:cme}, the estimator $\hat{\theta}_a^{\text{FB-IVW}}$ inherits the robustness and efficiency properties of $\hat{\theta}_a^{\text{FB-Xonly}}$, while further exploiting the prognostic information contained in $U$ through IVW. This construction is particularly well suited to settings with covariate mismatch across regions, as it avoids discarding auxiliary-region observations while efficiently incorporating additional target-region-specific covariates. From an alternative perspective, \eqref{eq:FB-IVW} can be viewed as an estimator plus a consistent estimator of zero, which is conceptually related to the augmentation framework of \cite{yang2020combining}; further details are provided in Appendix A.3 in the Supplementary Materials.

\vspace{-30pt}
\section{Conformal selective borrowing under outcome drift}\label{sec:CSB}

%violation
%target data enable us examine the violation
%in stead of all or nothing, partial hold for subset
%if we can identify this subset, we can use this subset for FB-IVW. this motivate us exmiane individually, then make decision collectively. thus two important parts: 1. conformal p-value 2. target tuning for selection threshold. then we can use this subset for FB-IVW and propose FRT in next section.

While Assumption~\ref{ass:cme} provides a basis for integrating auxiliary-region data into target-region treatment effect estimation, it may not hold in practice due to unmeasured regional differences, such as variation in healthcare systems or data collection protocols. As a result, full borrowing approaches that rely on Assumption~\ref{ass:cme} may yield biased estimation and invalid inference when conditional outcome incompatibility exists across regions.

Fortunately, because data from the target region are available as a benchmark, violations of Assumption~\ref{ass:cme} can be empirically assessed using the observed data \citep{Robertson2025}. When evidence suggests that Assumption~\ref{ass:cme} is violated, a conservative strategy is to discard all auxiliary-region data and rely solely on target-region information. However, this approach may incur substantial efficiency loss if Assumption~\ref{ass:cme} holds for a subset of auxiliary-region observations, as is the case in the motivating example (see Figure~\ref{fig:rd-sel}).

This motivates the goal of selectively identifying auxiliary-region observations exchangeable with the target region and borrowing information only from them.\rev{ We} pursue this goal in two steps. First, we leverage conformal p-values to assess exchangeability at the individual level for auxiliary-region observations. Second, we determine a selection threshold by minimizing the mean squared error (MSE) of a class of RSATE estimators, thereby making a collective borrowing decision based on the conformal p-values across auxiliary-region observations. \rev{Because the selection threshold is data-adaptive, in Section~\ref{FRT} we recommend conditional randomization inference that re-runs the selection procedure within each target-region re-randomization to preserve exact type~I error control.} We construct treatment effect estimators analogous to \eqref{eq:FB-IVW} using only data from the target region and the selected auxiliary-region observations. We refer to this procedure as \emph{Conformal Selective Borrowing (CSB)}, and provide detailed methodology in the following subsections.

\vspace{-10pt}
\subsection{Conformal p-value}

% we first give a general procedure for computing conformal p-value for one data point
% 1. data split 2. training and conformal score 3. cal and conformal p-value
% remark1: CQR, NN
% remark2: split, jackknife

In this subsection, we present the procedure for computing a conformal p-value for a single auxiliary-region observation $(X_j,Y_j)$ with $R_j=0$ and $A_j=a$. The goal is to assess its exchangeability with respect to the corresponding treatment group in the target region, that is, $\{(X_i,Y_i)\}_{i\in\mathcal{R}_a}$, where $\mathcal{R}_a=\{i:R_i=1,A_i=a\}$ denotes the index set of target-region patients receiving treatment $a$. The steps are as follows.

\vspace{-10pt}

\begin{enumerate}[label=\arabic*., leftmargin=*]

\item \textbf{Data splitting}: Randomly partition $\mathcal{R}_a$ into $K$ mutually exclusive folds, $\mathcal{R}_a=\bigcup_{k=1}^{K}\mathcal{R}_{a}^k$.

\item \textbf{Training}: For each fold $k$, fit a regression function $\hat{\mu}_a^{-k}({X})$ using data from the remaining $K-1$ folds, leaving out $\mathcal{R}_{a}^k$.

\item \textbf{Conformal score}: For the auxiliary-region observation $(X_j,Y_j)$, compute its conformal score relative to each target-region observation as
$$
s_j^{(i)}=\left|Y_j-\hat{\mu}_a^{-k(i)}({X}_j)\right|,\qquad i\in\mathcal{R}_a,
$$
where $k(i)\in\{1,\ldots,K\}$ denotes the fold index such that $i\in\mathcal{R}_{a}^{k(i)}$.

\item \textbf{Calibration}: For each target-region observation $(X_i,Y_i)$, compute the conformal score
$$
s_i=\left|Y_i-\hat{\mu}_a^{-k(i)}({X}_i)\right|,\qquad i\in\mathcal{R}_a.
$$

\item \textbf{CV+ p-value}: Define the CV+ conformal p-value for $(X_j,Y_j)$ as
$$
p_j=\frac{\sum_{i\in\mathcal{R}_a}\mathbb{I}\{s_i\ge s_j^{(i)}\}+1}{|\mathcal{R}_a|+1}.
$$
\end{enumerate}

Intuitively, if $(X_j,Y_j)$ is not exchangeable with the target-region observations $\{(X_i,Y_i)\}_{i\in\mathcal{R}_a}$, its conformal scores $\{s_j^{(i)}\}_{i\in\mathcal{R}_a}$ will tend to be large relative to $\{s_i\}_{i\in\mathcal{R}_a}$, which are computed from observations from the target region. Consequently, the p-value $p_j$ will be small, providing evidence against exchangeability and indicating that $(X_j,Y_j)$ should not be borrowed.

\vspace{-20pt}

\begin{remark}[Conformal score]
Although we use absolute residuals as the conformal score for illustration, the above procedure applies more generally to any conformal score that quantifies the similarity between $(X_j,Y_j)$ and the target-region observations. For example, the conformal quantile regression (CQR) score \citep{romano2019} is based on fitted conditional quantiles and nearest-neighbor-based scores can be used for binary outcomes \citep{shafer2008tutorial}. As a result, the proposed procedure is model-free and broadly applicable to modern information borrowing settings.
\end{remark}

\vspace{-20pt}

\begin{remark}[Split conformal p-value]
A simpler conformal p-value can be obtained using split conformal methods \citep{papadopoulos2002inductive}, which rely on a single partition of the data into training and calibration sets. We adopt the CV+ approach \citep{barber2021} because the target-region sample $\mathcal{R}_a$ is typically limited in size, and CV+ allows more efficient use of the available data by leveraging all observations for both training and calibration. Preliminary simulation results suggest that CV+ yields improved finite-sample selection performance compared with split conformal p-values.    
\end{remark}

\vspace{-10pt}
\subsection{MSE-guided selection threshold}

Applying the procedure in the previous subsection yields conformal p-values for all auxiliary-region observations. While larger values of $p_j$ indicate greater compatibility with the target region, a selection threshold $\gamma_a\in[0,1]$ is required to determine the subset of auxiliary-region observations to be borrowed for estimating $\theta_a$. Specifically, for treatment arm $a$, we define the selected set
$
\mathcal{E}_a(\gamma_a)=\{j:R_j=0,A_j=a,p_j\ge\gamma_a\}.
$

One possible approach is to frame this as a multiple testing problem and choose $\gamma_a$ to control the false discovery rate \citep{jin2023selection}. However, due to the limited sample size in the target region, such $\gamma_a$ may lead to low power to detect incompatible auxiliary-region observations and may consequently induce substantial bias in the resulting RSATE estimator.
Instead, we view the conformal p-values as a similarity ranking and determine $\gamma_a$ to minimize the MSE of the RSATE estimator introduced below. This perspective is closely related to targeted tuning of causal estimators \citep{rothenhausler2024model,Zhu2024}.

For a given threshold $\gamma_a$, we construct an RSATE estimator analogous to \eqref{eq:FB-IVW} using data from the target region and the selected auxiliary-region observations,
$$
\hat{\theta}_a^{\text{CSB-IVW}}(\gamma_a)
=\frac{1}{n_{\mathcal{R}}}\sum_{i=1}^n\left[
R_i\,\hat{Y}_i^{\text{CSB}}
+\hat{\pi}^{\text{CSB}}(X_i)\frac{\mathbb{I}(A_i=a,p_i\ge\gamma_a)}{
\hat{e}^{\text{CSB}}_a(X_i)
}
\{Y_i-\hat{Y}_i^{\text{CSB}}\}
\right],
$$
where we set $p_i\equiv1$ for all target-region observations with $R_i=1$. The quantities $\hat{\pi}^{\text{CSB}}$ and $\hat{Y}_i^{\text{CSB}}$ are defined analogously to $\hat{\pi}^{\text{FB}}$ and $\hat{Y}_i^{\text{FB}}$, except that the full set of auxiliary-region observations $\{(X_j,Y_j)\}_{j:R_j=0,A_j=a}$ is replaced by the selectively borrowed subset $\{(X_j,Y_j)\}_{j\in\mathcal{E}_a(\gamma_a)}$. In addition, $\hat{e}^{\text{CSB}}_a(X_i)$ denotes an estimator of the conditional probability $\mathbb{P}[A_i=a,\,p_i\ge\gamma_a\mid X_i]$. For notational simplicity, we suppress the dependence of these quantities on $\gamma_a$.

This formulation reveals that $\hat{\theta}_a^{\text{CSB-IVW}}(\gamma_a)$ interpolates between the full-borrowing and no-borrowing estimators. When $\gamma_a=0$, $\mathcal{E}_a(0)=\{j:R_j=0,A_j=a\}$ and $\hat{\theta}_a^{\text{CSB-IVW}}(0)$ coincides with the full-borrowing estimator $\hat{\theta}_a^{\text{FB-IVW}}$. When $\gamma_a=1$, $\mathcal{E}_a(1)=\emptyset$ and $\hat{\theta}_a^{\text{CSB-IVW}}(1)$ reduces to the no-borrowing estimator $\hat{\theta}_a^{\text{NB-AllCov}}$. For intermediate values $0<\gamma_a<1$, $\hat{\theta}_a^{\text{CSB-IVW}}(\gamma_a)$ selectively borrows information from the most compatible auxiliary-region observations.

We decompose
$
\text{MSE}_a(\gamma_a)
=[\mathbb{E}\{\hat{\theta}_a^{\text{CSB-IVW}}(\gamma_a)\}-\theta_a]^2
+\mathbb{V}\{\hat{\theta}_a^{\text{CSB-IVW}}(\gamma_a)\}.
$
Since $\theta_a$ is unknown, we approximate the squared bias using the consistent no-borrowing estimator $\hat{\theta}_a^{\text{NB-AllCov}}$ as a benchmark,
$
[\mathbb{E}\{\hat{\theta}_a^{\text{CSB-IVW}}(\gamma_a)\}-\theta_a]^2
\approx
\mathbb{E}\{\hat{\theta}_a^{\text{CSB-IVW}}(\gamma_a)-\hat{\theta}_a^{\text{NB-AllCov}}\}^2
-\mathbb{V}\{\hat{\theta}_a^{\text{CSB-IVW}}(\gamma_a)-\hat{\theta}_a^{\text{NB-AllCov}}\}.
$
Variance terms can be approximated via the nonparametric bootstrap, yielding an estimated $\widehat{\text{MSE}}_a(\gamma_a)$. We evaluate $\widehat{\text{MSE}}_a(\gamma_a)$ over a grid $\gamma_a\in\mathcal{G}\subset[0,1]$ and select $\hat{\gamma}_a=\arg\min_{\gamma_a\in\mathcal{G}}\widehat{\text{MSE}}_a(\gamma_a)$. The final estimator is
$
\hat{\tau}^{\text{CSB-IVW}}
=\hat{\theta}_1^{\text{CSB-IVW}}(\hat{\gamma}_1)
-\hat{\theta}_0^{\text{CSB-IVW}}(\hat{\gamma}_0),
$
and this MSE-guided threshold selection strategy \rev{is a prespecified data-adaptive rule} and is designed to improve finite-sample performance relative to a fixed choice of $\gamma_a$.
\rev{Because $\hat{\gamma}_a$ is selected from the observed data, we recommend the conditional Fisher randomization test in Section~\ref{FRT} for hypothesis testing to account for selection uncertainty.}
Since both the FB and NB estimators are included as special cases, the resulting estimator is expected to improve upon both. The detailed algorithm is provided in Appendix B in the Supplementary Materials.

\vspace{-30pt}
\section{Fisher randomization test for type I error rate control}\label{FRT}

%asymptotic 1. doubly robust 2. selection consistency
%to protect inference validity, we provide a finite-sample exact, model-free, post-selection valid FRT.

%procedure: 1. sharp null on target region 2. specify a test stat 3. conditional on A of auxiliry region, permute A of target region and compute p-value

% remark1: theorem show validity only rely on randomization, for any test stat.
% remark2: MC
% remark3: one sided

% theorem

After obtaining the RSATE estimator $\hat{\tau}^{\text{CSB-IVW}}$, asymptotic inference is available under the following conditions: (i) either the outcome regression model or the propensity score model is correctly specified, and (ii) the selection of auxiliary-region observations satisfying Assumption~\ref{ass:cme} is consistent. To further safeguard inferential validity, we additionally propose a finite-sample exact, model-free, post-selection-valid Fisher randomization test (FRT), which controls the type I error rate even when these assumptions fail. The procedure is as follows.
\vspace{-10pt}
\begin{enumerate}[label=\arabic*., leftmargin=*]
\item \textbf{Sharp null in the target region}: Under the sharp null hypothesis
$H_0: Y_i(0)=Y_i(1)$ for all $i\in\mathcal{R}=\{i:R_i=1\}$,
all missing potential outcomes in the target region can be imputed using the observed outcomes $Y_i$.
\item \textbf{Test statistic}: Specify a test statistic and compute its observed value $T(\boldsymbol{A})$ based on the observed assignment vector $\boldsymbol{A}=(A_1,\ldots,A_n)$.
\item \textbf{Conditional randomization}: Fix the treatment assignments for auxiliary-region observations at their observed values, that is, set $A_i^*=A_i$ for all $i$ with $R_i=0$. Randomize the treatment assignments $A_i^*$ for target-region observations according to the original randomization design, yielding a reassigned vector $\boldsymbol{A}^*=(A_1^*,\ldots,A_n^*)$.
\item \textbf{P-value}: Repeatedly generate $\boldsymbol{A}^*$ and compute the corresponding test statistics $T(\boldsymbol{A}^*)$. The FRT p-value is defined as
$
p^{\text{FRT}}
=\mathbb{P}_{\boldsymbol{A}^*}\!\left\{\,|T(\boldsymbol{A}^*)|\geq |T(\boldsymbol{A})|\,\right\},
$
where the probability is taken with respect to the conditional randomization distribution.
\end{enumerate}
\vspace{-20pt}
\begin{remark}[Test statistic]
The FRT is valid for any choice of test statistic. Consequently, any of the estimators introduced in this paper can be used as a test statistic, including $\hat{\tau}^{\text{FB-IVW}}$ even when Assumption~\ref{ass:cme} fails, since the FRT correctly reproduces its finite-sample distribution under the sharp null. However, when biased auxiliary-region observations are borrowed, the power of the test may be reduced relative to an FRT based on $\hat{\tau}^{\text{NB-AllCov}}$. We therefore recommend using $\hat{\tau}^{\text{CSB-IVW}}$ as the test statistic, as it selectively borrows compatible auxiliary-region information and can improve power. A key point is that, when computing $T(\boldsymbol{A}^*)$ under randomization, the CSB procedure must be allowed to reselect auxiliary-region observations for each realization of $\boldsymbol{A}^*$ in order to properly account for selection uncertainty.
\end{remark}
\vspace{-20pt}
\begin{remark}[Monte Carlo approximation]
In practice, the randomization distribution can be approximated via Monte Carlo sampling of $\boldsymbol{A}^*$ rather than enumerating all possible treatment assignments.
\end{remark}
\vspace{-20pt}
\begin{remark}[One-sided hypothesis and randomization-based interval]
The above discussion focuses on two-sided testing. For a one-sided hypothesis, such as $H_0: Y_i(0)>Y_i(1), \forall i\in\mathcal{R}$, the FRT p-value is defined as
$
p^{\text{FRT}}
=\mathbb{P}_{\boldsymbol{A}^*}\!\left\{T(\boldsymbol{A}^*)\ge T(\boldsymbol{A})\right\}.
$
The randomization-based interval can be obtained by inverting the FRT \citep{luo2021leveraging,zhu2023pair,zhu2024rejoinder}.
\end{remark}

\vspace{-10pt}
The following theorem establishes the validity of the FRT; its proof is provided in Appendix C in the Supplementary Materials.
\vspace{-10pt}
\begin{theorem}
Under the null hypothesis $H_0$, for any $\alpha\in(0,1)$,
$
\mathbb{P}_{\boldsymbol{A}}\!\left(p^{\text{FRT}}\le \alpha \mid \boldsymbol{A}_{\mathcal{E}}\right)\le \alpha,
$
where $\mathbb{P}_{\boldsymbol{A}}$ denotes probability with respect to the randomization distribution of the assignment vector $\boldsymbol{A}$, treating the potential outcomes and covariates as fixed, and $\boldsymbol{A}_{\mathcal{E}}$ denotes the subassignment vector corresponding to auxiliary-region observations. Moreover,
$$
\mathbb{P}_{\boldsymbol{A}}\!\left(p^{\text{FRT}}\le \alpha\right)
=\sum_{\boldsymbol{A}_{\mathcal{E}}}
\mathbb{P}_{\boldsymbol{A}}\!\left(p^{\text{FRT}}\le \alpha \mid \boldsymbol{A}_{\mathcal{E}}\right)
\mathbb{P}(\boldsymbol{A}_{\mathcal{E}})
\le \alpha \sum_{\boldsymbol{A}_{\mathcal{E}}}\mathbb{P}(\boldsymbol{A}_{\mathcal{E}})
=\alpha.
$$
\end{theorem}

\vspace{-30pt}
\section{Simulation studies} \label{Simulation}

\vspace{-10pt}
\subsection{Simulation design}
To evaluate the finite sample performance of the proposed method and examine how covariate informativeness affects performance, we conduct simulation studies with varying signal strengths of $X$ under both scenarios where (i) $X = (X_1, X_2)$ and $U$  were generated independently, and (ii) $X$ and $U$ were jointly generated with a moderate correlation. Individuals are independently assigned to different regions with the sampling indicator $R \sim \text{Bernoulli}(\pi(X))$. From a large superpopulation,
$\{n_{\mathcal{R}},n_{\mathcal{E}}\}=\{600,1000\}$  were sampled for the final study. Treatments are assigned as $A \sim \text{Bernoulli}(0.5)$. For samples in the target region, the potential outcomes are generated by $Y(a) \sim \mathcal{N}(\beta_{a0}+X^\top \beta_{a1}+U \alpha_{a},1)$. For samples in the auxiliary region, a random proportion $\rho=50\%$ of the samples is biased in arm $a$ with a hidden bias $b_a$. The potential outcomes for biased samples are generated by $Y(a) \sim \mathcal{N}(-b_a+\beta_{a0}+X^\top \beta_{a1},\epsilon)$, where $\epsilon=\{0.1, 0.5, 1, 1.5\}$, $b_0=6$, and $b_1=10$. The remaining samples in the auxiliary region remain unbiased with $Y(a) \sim \mathcal{N}(\beta_{a0}+X^\top \beta_{a1}+U \alpha_{a},\epsilon)$. The detailed data-generating process is summarized in Appendix D.1 in the Supplementary Materials.

We quantify signal strength using a conditional \( R^2 \)-based signal-to-noise metric which is defined as \( R^2_{X|U} / R^2_{U|X} \), where \( R^2_{X|U}=\text{SSE}(X|U)/\text{SSE}(U) \) denotes the proportion of variance in the outcome explained by \( X \) after conditioning on \( U \), and \( R^2_{U|X}=\text{SSE}(U|X)/\text{SSE}(X) \) captures the remaining variance explained by \( U \) after accounting for \( X \). A higher signal ratio indicates stronger predictive power of \rev{observed} covariates relative to unmeasured variables.

%Our simulation design enables control over signal strength through two key mechanisms. First, the coefficient \( \alpha_a \) in the outcome model directly modulates the contribution of \( U \) to potential outcomes with larger values of \( \alpha_a \) imply stronger unmeasured confounding. By varying \( \alpha_0 \) and \( \alpha_1 \), we generate outcome scenarios with different levels of dependence on \( U \). Second, the precision of the outcome model is governed by the error variance parameter \( \epsilon \), where smaller \( \epsilon \) implies higher signal-to-noise ratio for the covariates. Together, these parameters allow us to simulate a range of signal conditions from settings where \( X \) captures most of the predictive information to those dominated by \( U \). 

We evaluated the performance of FRTs in terms of type I error control and statistical power. We considered scenarios in which $\epsilon = 0.5$ and the hidden bias was introduced only in the control arm, for simplicity. To assess type I error, we generated data under the null hypothesis of no regional treatment effect and examined the empirical rejection rates across increasing magnitudes of hidden bias $b_0=2,4,6,8$. To assess power, we simulated data under a range of nonzero true treatment effects, varying the level of hidden bias in the control arm. 

For method comparison, we evaluate six estimators under varying borrowing strategies and covariate adjustments. Specifically, we consider three borrowing schemes: NB, FB and CSB with 10 folds CV and absolute residual conformal scores. For each borrowing scheme, we assess two versions: one that adjusts for observed covariates \( X \) (X-only), and one that adjusts for both \( X \) and \( U \) (AllCov/IVW). We replicate the simulation 500 times per scenario. Additional simulation results are provided in Appendix D.2 in the Supplementary Materials.

\vspace{-10pt}
\subsection{Simulation results}
\begin{figure}[htbp]
    \centering \includegraphics[width=\textwidth]
    {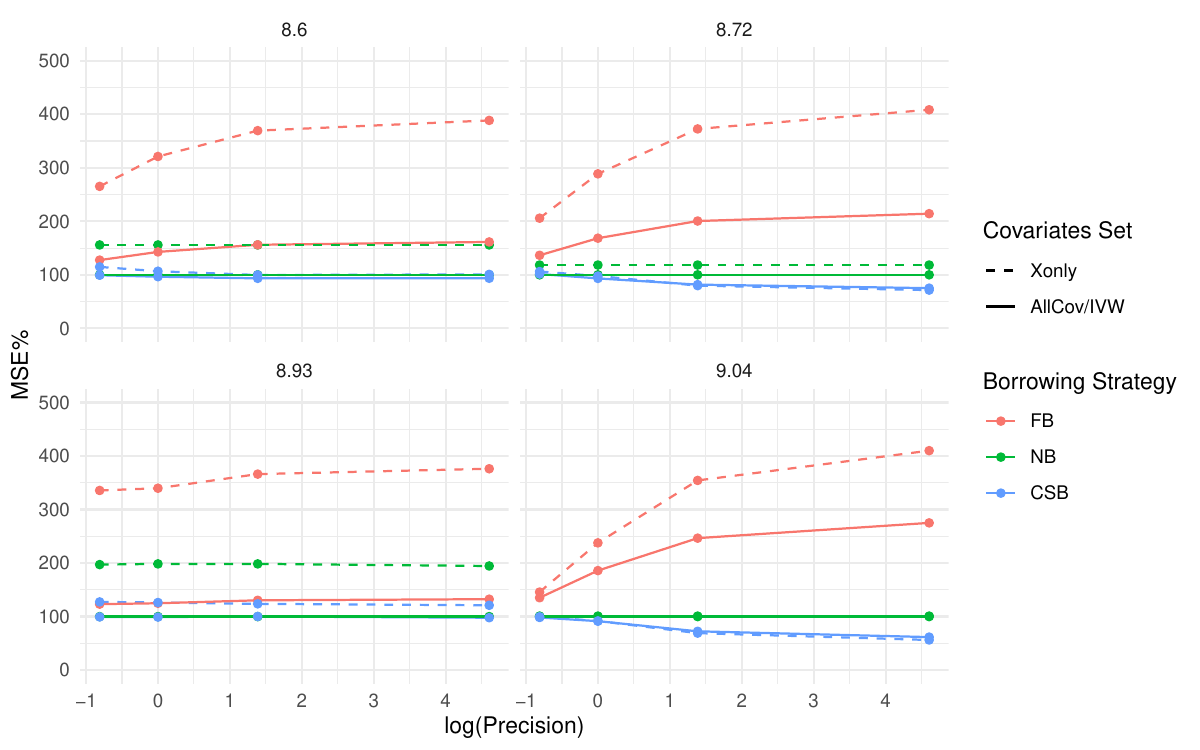}
    \caption{MSE\% vs log(Precision) across different signal of $X$ with correlated $X$ and $U$.}
    \label{figure2}
\end{figure}

Figures~\ref{figure2} present the results under varying levels of precision and signal strength of $X$, which are reported as mean squared error percentages (MSE\%) of the treatment effect estimator $\hat{\tau}$, standardized relative to the NB-AllCov estimator. We observe that MSE\% decreases as either the signal strength of $X$ or precision increases. This pattern reflects more accurate treatment effect estimation when observed covariates are informative and outcome noise is reduced. The CSB estimators consistently outperform the NB baseline across all settings, achieving MSE\% reductions of approximately 10\% to over 50\%, with the greatest improvements seen in high-precision, high-signal settings, while FB estimators perform the worst since they include all biased samples in estimation. When separating biased samples is difficult, CSB is at least as good as NB, underscoring its robustness to borrowing-related bias. We also observe significant performance improvements when comparing AllCov/IVW estimators with X-only estimators. This illustrates the advantage of incorporating additional covariate information and, notably, for the X-only estimators, even without access to $U$, using $X$ yields more accurate estimates in terms of MSE\% when $X$ is moderately correlated with the unmeasured covariate. This makes intuitive sense as when correlation exists, $X$ can partially capture the variation in $U$. 
\vspace{-35pt} 

\vspace{-10pt} 

\begin{figure}[htbp]
    \centering   \includegraphics[width=\textwidth]
    {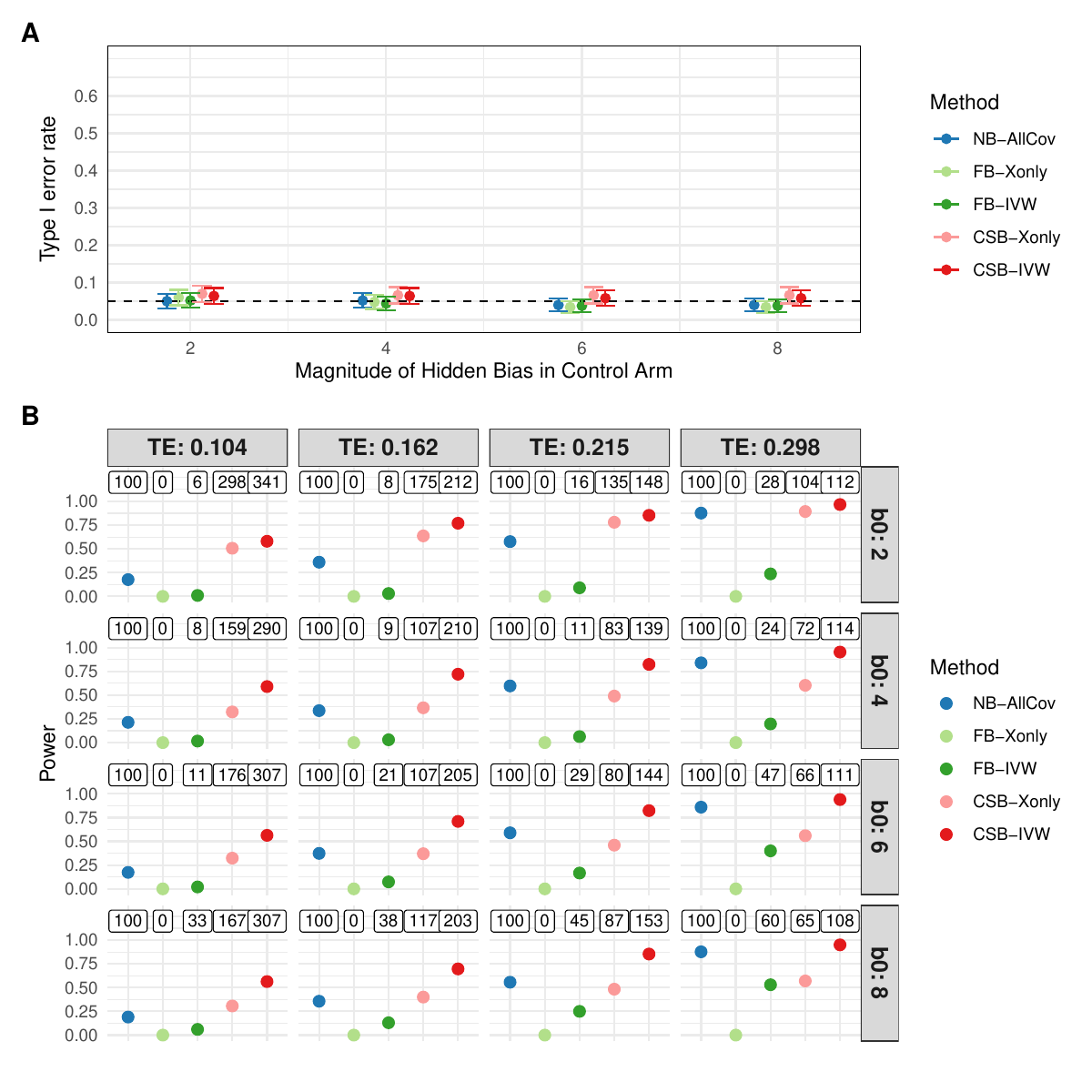}
    \caption{Type I error rates of FRTs under varying magnitudes of hidden bias and statistical power under varying true treatment effects and hidden bias.}
    \label{figure3}
\end{figure}

Figure \ref{figure3} presents the empirical type I error rates across increasing levels of bias and summarizes the power performance under a range of true treatment effects. Across all scenarios, the FRTs successfully maintained nominal type I error rate control at the 0.05 level. Power comparisons reveal a clear and consistent hierarchy among the three borrowing strategies. The NB-AllCov estimator provides a safe baseline but at the cost of lower power. The FB approach performs worst, particularly under outcome heterogeneity, due to indiscriminate pooling of biased auxiliary samples. The CSB estimators consistently achieve the highest power across all settings as they selectively integrate only outcome-compatible data. Across all borrowing schemes, the AllCov/IVW estimators achieve higher statistical power than X-only estimators. The improvement is particularly evident for the CSB-IVW estimator, which consistently performs the best overall.

%Taken together, these results show that CSB-IVW consistently achieves the lowest MSE across all simulation settings, outperforming NB and FB estimators. Overall, CSB-IVW provides the most reliable balance between robustness and efficiency, and achieves the highest power, while controlling type I error rate.

\vspace{-15pt} 
%\vspace{-20pt} 
%\begin{figure}[htbp]
%    \centering   \includegraphics[width=\textwidth]
%    {Picture4.pdf}
%    \caption{Statistical power under varying true treatment effects and hidden bias.}
 %   \label{figure4}
%\end{figure}

\vspace{-30pt}
\section{Real data illustration} \label{Realdata}
\vspace{-10pt}
\subsection{Data preparation}
We apply the proposed methods to a Phase 3, international, multi-center, placebo-controlled clinical trial, known as the POWER trial. For illustrative purposes, the regulatory question of interest we address is whether enobosarm is effective for the prevention and treatment of muscle wasting in patients with advanced non-small cell lung cancer in North America. In North America ($R=1$), 39 patients were randomized to receive enobosarm 3 mg ($A=1$) and 30 patients were randomized to receive placebo ($A=0$). The trial also included 576 patients from different regions including Europe and South America ($R=0$) which can serve as auxiliary data to improve North American treatment effect estimation and inference. 

We used the percentage change in lean body mass (LBM) at Day 84 as our primary endpoint. There are 7 shared baseline covariates, including age, sex, ECOG status, chemotherapy use, cancer stage, histology, and baseline lean body mass. These covariates capture key baseline characteristics in patient profiles and serve as the basis for addressing covariate incomparability. In addition, the weight-loss flag, which is an important prognostic factor indicating whether body weight decreased by 5\% or more in the 6 months prior to the study, is available in the North American data but is missing in the auxiliary regions. 

 A 1:4 nearest-neighbor matching procedure is implemented to enhance covariate comparability. Observations in auxiliary regions with missing values or covariates outside the support of the target region were removed before matching. The matching was performed using shared covariates, which retain all target region samples and $n_{\mathcal{E}}=276$ patients from the auxiliary regions including 155 and 121 patients in the control and treatment arm respectively. More details about the real data are available in Appendix E in the Supplementary Materials.

\vspace{-10pt}
\subsection{Data analysis results}
We implement NB, FB, and CSB with absolute residual score, each with two covariate adjusting schemes (X-only and AllCov/IVW) to estimate the RSATE and conduct FRTs. We also consider the simple difference-in-means estimator (DiM) as the benchmark. In addition, we compute asymptotic standard errors, confidence intervals, and p-values for all methods.

\begin{figure}[htbp]
    \centering   \includegraphics[width=\textwidth]
    {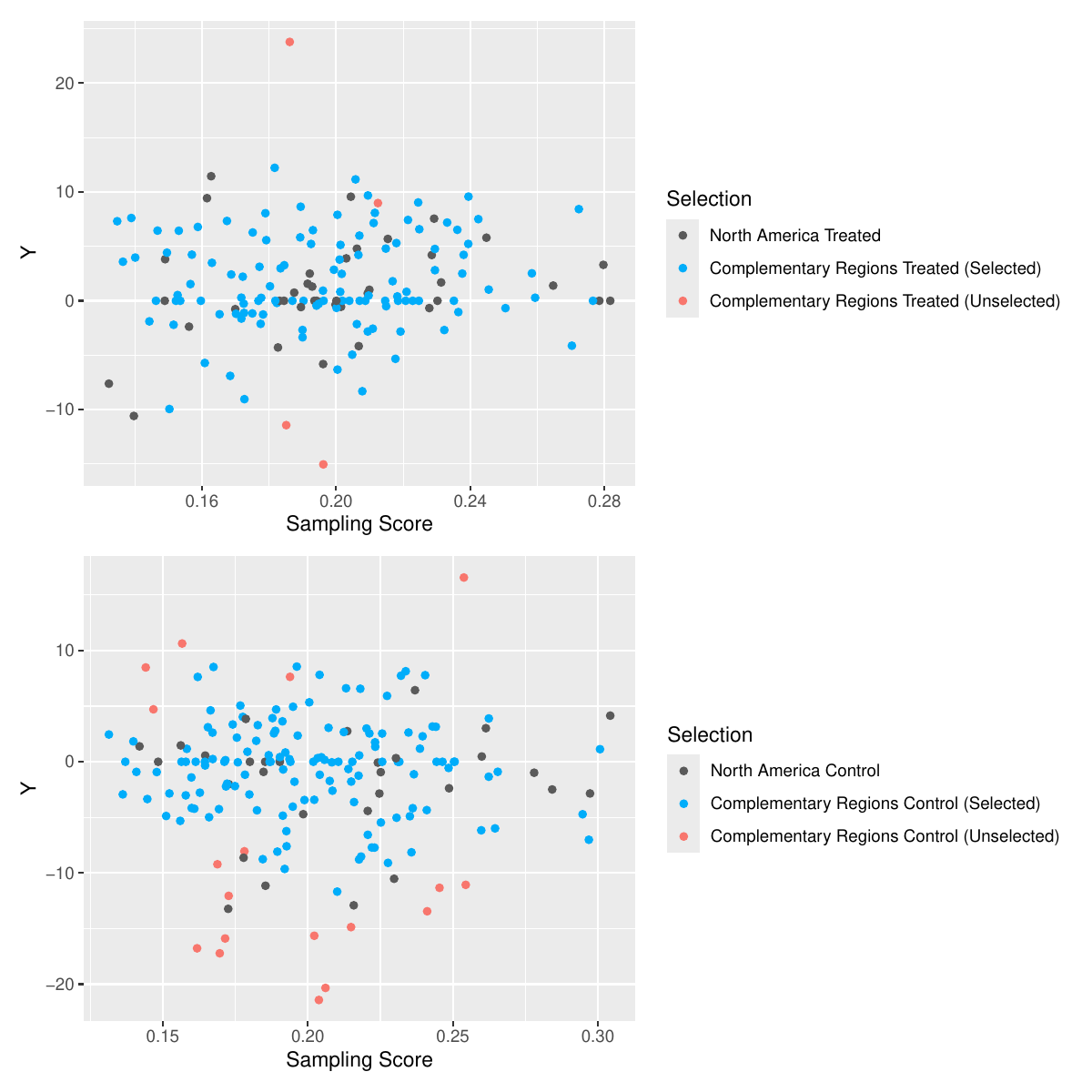}
    \caption{Outcome (LBM on Day 84) VS. Sampling score distribution}
    \label{fig:rd-sel}
\end{figure}

\begin{figure}[h]
    \centering   \includegraphics[width=\textwidth]
    {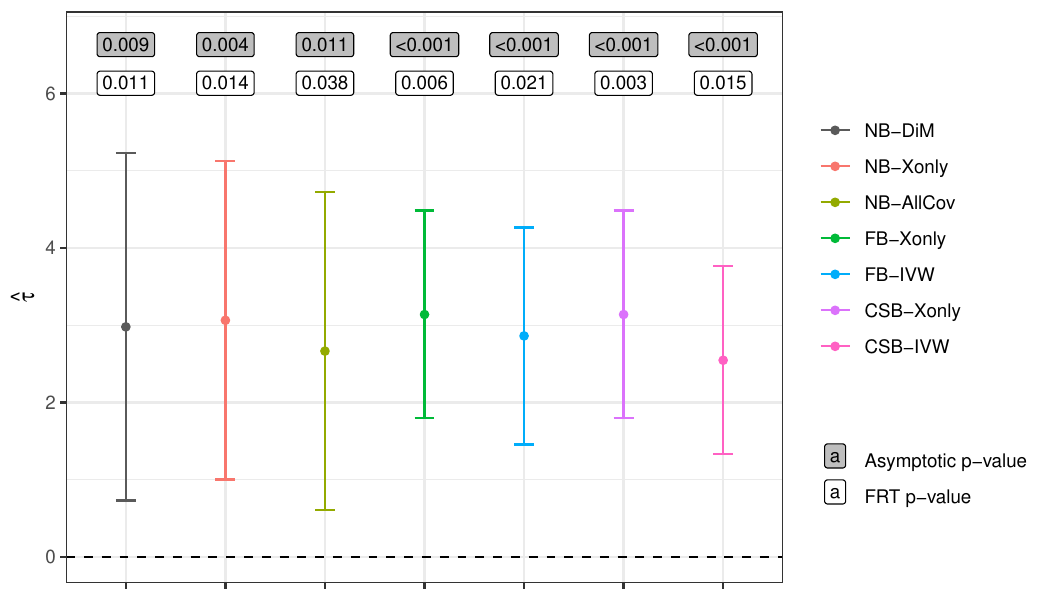}
    \caption{Comparison of estimated treatment effects and p-values across borrowing strategies.\rev{ The upper (lower) numbers report asymptotic (FRT) p-values.}}
    \label{figure7}
\end{figure}

Figure \ref{fig:rd-sel} displays the samples selected from the auxiliary regions under the CSB-IVW. Most of the selected auxiliary samples align closely with the North American samples in both sampling scores and outcomes, suggesting that conformal selection effectively identifies exchangeable auxiliary data. In contrast, the unselected auxiliary samples tend to deviate from the North American samples, particularly at the extreme ends of the sampling score distribution, indicating limited overlap in covariate space. This pattern demonstrates that the CSB-IVW procedure successfully prioritizes auxiliary data that are more similar to the target region, thereby reducing hidden bias from potential outliers.
\vspace{-10pt} 
\vspace{-10pt} 
Figure \ref{figure7} presents the estimated RSATE, 95\% confidence intervals, and asymptotic and FRT p-values under different borrowing strategies. The X-only estimators tend to overestimate the treatment effect, likely because they fail to adjust for covariate misalignment. In contrast, the AllCov/IVW estimators mitigate this inflation by accounting for information from the additional covariate. The CSB-IVW estimate is slightly lower than both the NB-AllCov and FB-IVW estimates because it selectively borrows from both treatment arms and excludes more control-arm patients with lower outcomes, thereby adjusting the estimate downward. In addition, compared with both NB and FB, CSB-IVW improves precision substantially, reducing the width of the confidence interval by roughly 10-45\%. Across both asymptotic inference and FRTs, CSB-IVW produces consistently smaller p-values than its no-borrowing and full-borrowing counterparts. Although asymptotic p-values tend to be smaller than their FRT counterparts, reflecting the conservativeness of randomization-based inference, the agreement between the two types of p-values under CSB-IVW supports the reliability of the estimated treatment effect. The estimated LBM gain of approximately 2.5\% corresponds to a clinically meaningful improvement in muscle mass recovery, suggesting stronger and more stable evidence for a positive RSATE.
\vspace{-10pt} 

\vspace{-50pt}
\section{Discussion} \label{Discussion}
In this article, we framed multi-regional trial analysis around a formal estimand for the region-specific average treatment effect (RSATE) in a prespecified target region and developed a selective information borrowing framework for its estimation and inference. By combining an inverse variance weighting estimator with conformal subset selection and conditional randomization tests, our CSB-IVW procedures improve efficiency and power for RSATE while preserving valid type I error control under covariate shift, covariate mismatch, and outcome drift. These properties make the resulting RSATE inferences particularly well suited for local regulatory assessment, where clearly defined estimands and transparent borrowing rules are increasingly required. 

Our framework relies on standard causal identification assumptions within the target region (consistency, randomization, and positivity) and on the availability of covariates that are sufficiently predictive to distinguish comparable auxiliary-region patients. When regions are nearly non-overlapping in covariates or when outcome drift is pervasive, the proposed procedures naturally shrink toward the target-only estimator, providing conservative yet transparent borrowing for regulatory decision-making.

Although we focus on the MRCT setting, our methodology extends naturally to augmented randomized clinical trials that incorporate both external controls and external treatments, while accommodating external data with mismatched covariates, thereby extending the methods in \cite{Zhu2024} to broader practical scenarios and aligning with current regulatory interest in using real-world data for confirmatory analyses \citep{FDA2023_RealWorldDataRegistries}.
Extending CSB to time-to-event outcomes is feasible \citep{Candes2023} and particularly useful in oncology trials where censoring and event-time patterns differ across regions. The conditional randomization test also generalizes to these settings \citep{heng2025design}. 
%The methodology further applies beyond MRCTs, including small randomized trials augmented with external treatment and control arms, aligning with current regulatory interest in using real-world data for confirmatory analyses \citep{FDA2023_RealWorldDataRegistries}.
Related to the covariate-mismatch issue, important prognostic factors, such as ECOG performance status, may be inconsistently recorded across data sources. Proxy measures such as the Charlson comorbidity index can partially capture these latent factors. In such cases, proximal causal methods \citep{Tchetgen2024} provide a potential path for robust information borrowing when key confounders are not directly observed.

\backmatter

\vspace{-30pt}
\section*{Acknowledgements}
\vspace{-10pt}
This project is supported by the Food and Drug Administration (FDA) of the U.S. Department of Health and Human Services (HHS) as part of a financial assistance award, U01FD007934, totaling \$2,556,429 over three years, funded by the FDA/HHS. This work is also supported by R01AG066883, funded by the NIH/HHS. The contents are those of the authors and do not necessarily represent the official views of, nor an endorsement by, FDA/HHS, NIH/HHS, or the U.S. Government.

\vspace{-30pt}
\section*{Supplementary Materials}
\vspace{-10pt}
Web Appendices A–E include technical details, algorithms, proofs, simulations, and supplementary data analyses. R code is available at \url{github.com/chenxi0217/RSATE-in-MRCT}.

\vspace{-30pt}
\section*{Data Availability}
The data are not publicly available due to participant privacy concerns.

\vspace{-30pt}
%\begin{singlespace}
%\small
\bibliographystyle{biom}
\bibliography{_ref}
%\end{singlespace}

\label{lastpage}

\end{document}

\date{}

\pagerange{\pageref{firstpage}--\pageref{lastpage}} 
\volume{}
\pubyear{}
\artmonth{}

\doi{}

\label{firstpage}

\maketitle

Appendix A presents the asymptotic variance estimators and inference procedures for the estimation approaches introduced in the main text. Appendix B outlines the algorithm used to adaptively select the threshold. Appendix C gives a proof of Theorem 1 on the validity of the Fisher randomization test. Appendix D summarizes the data-generating mechanisms and parameters used in the simulation studies, and additional simulation results. Appendix E reports supplementary results for the real-data application. Appendix F extends FB-IVW to multiple auxiliary regions with different shared covariates by using region-specific identification and optimally combining region-wise estimates to infer the target-region treatment effect.

\section*{Appendix A: Semiparametric efficient estimators and asymptotic inference} \label{inference}

Let $z_{1 - \alpha/2}$ denote the \rev{$(1-\alpha/2)$ quantile} of the standard normal distribution and $\pi_\mathcal{R}=\mathbb{E}\{\pi(X)\}$ denote the sample ratio of target region data. 
Let $O=(R,A,Y,X,U)$ denote the \rev{observed} data, where $U$ is partially missing for $R=0$.

\subsection*{A.1 Target-only analysis (NB-AllCov)}

% Recall that
% \begin{equation*}
% \hat{\tau}^{\text{NB-AllCov}}=\hat{\theta}_1^{\text{NB-AllCov}}-\hat{\theta}_0^{\text{NB-AllCov}}
% \end{equation*}
Let 
$$\xi_a(O) = \mu_a(X,U)
+\frac{\mathbb{I}(R=1,A=a)}{e_{a\mid1}(X)}
\{Y-\mu_a(X,U)\}
$$ and $\hat{\xi}(O_i)$ denote the empirical version of $\xi_a(O)$ with estimated nuisance function, that is,
\begin{equation*}
\hat{\xi}_a(O_i)=\hat{\mu}_a^{\text{NB}}(X_i,U_i)
+\frac{\mathbb{I}(R_i=1,A_i=a)}{e_{a\mid1}(X_i)}
\{Y_i-\hat{\mu}_a^{\text{NB}}(X_i,U_i)\}.
\end{equation*}
% The target-only EIF of $\tau$ is
% \begin{align*}
% \mathbb{IF}(\tau) &= \mathbb{IF}(\theta_1 - \theta_0) = \mathbb{IF}(\theta_1) - \mathbb{IF}(\theta_0) \\
% &= \frac{R}{\pi_{\mathcal{R}}} \left\{ \xi_1(O) - \xi_0(O) - \tau \right\}.
% \end{align*}
The variance estimator for $\sqrt{n} \hat{\tau}^{\text{NB-AllCov}}$ is
\begin{align*}
\hat{V}^{\text{NB-AllCov}} &= \frac{1}{n} \sum_{i=1}^{n} \left[ \frac{R_i}{\pi_{\mathcal{R}}} \left\{ \hat{\xi}_1(O_i) - \hat{\xi}_0(O_i) - \hat{\tau}^{\text{NB-AllCov}} \right\} \right]^2 \\
&= \frac{1}{n {\pi_{\mathcal{R}}}^2} \sum_{i:R_i=1} \left\{ \hat{\xi}_1(O_i) - \hat{\xi}_0(O_i) - \hat{\tau}^{\text{NB-AllCov}} \right\}^2.
\end{align*}
The asymptotic confidence interval is
$$
\left[
\hat{\tau}^{\text{NB-AllCov}} - z_{1 - \alpha/2} \sqrt{\hat{V}^{\text{NB-AllCov}}/{n}}, \ 
\hat{\tau}^{\text{NB-AllCov}} + z_{1 - \alpha/2} \sqrt{\hat{V}^{\text{NB-AllCov}}/{n}}
\right].
$$

% For a sanity check, we see that the variance estimator for $\hat{\tau}$ is
% \begin{align*}
% \hat{V} / n &= \frac{1}{n^2 {\pi_{\mathcal{R}}}^2} \sum_{i:R_i=1} \left\{ \hat{\xi}_1(O_i) - \hat{\xi}_0(O_i) - \hat{\tau}^{\text{NB-AllCov}} \right\}^2 \\
% &= \frac{1}{{n_{\mathcal{R}}}^2} \sum_{i:R_i=1} \left\{ \hat{\xi}_1(O_i) - \hat{\xi}_0(O_i) - \hat{\tau}^{\text{NB-AllCov}} \right\}^2.
% \end{align*}

\subsection*{A.2 Borrowing from auxiliary regions with shared covariates (FB-Xonly)}

% Recall that
% \begin{equation*}
% \hat{\tau}^{\text{FB-Xonly}}=\hat{\theta}_1^{\text{FB-Xonly}}-\hat{\theta}_0^{\text{FB-Xonly}}.
% \end{equation*}
Let 
$$\phi_a(O)=\frac{R}{\pi_\mathcal{R}}\mu_{a}(X)+\frac{\pi(X)}{\pi_\mathcal{R}}\frac{\mathbb{I}(A=a)}{e_a(X)}\{Y - \mu_{a}(X)\},
$$
and $\hat{\phi}_a(O_i)$ denote the empirical version of $\phi_a(O)$ with estimated nuisance functions, that is, 
\begin{equation*}
\hat{\phi}_a(O_i)=\frac{R_i}{\pi_\mathcal{R}}\hat{\mu}^{\text{FB}}_a(X_i)+\frac{\hat{\pi}^{\text{FB}}(X_i)}{\pi_\mathcal{R}}\frac{\mathbb{I}(A_i=a)}{e_a(X_i)}
\{Y_i-\hat{\mu}^{\text{FB}}_a(X_i)\}.\end{equation*}
% The EIF of $\tau$ is
% \[
% \mathbb{IF}(\tau) 
% = \mathbb{IF}(\theta_1 - \theta_0) 
% = \mathbb{IF}(\theta_1) - \mathbb{IF}(\theta_0) 
% = \phi_1(O) - \phi_0(O) - \frac{R}{\pi_R}\hat{\tau}^{\text{FB-Xonly}}.
% \]
The variance estimator for $\sqrt{n}\hat{\tau}^{\text{FB-Xonly}}$ is
\[
\hat{V}^{\text{FB-Xonly}} = \frac{1}{n}\sum_{i=1}^n \left\{\hat{\phi}_1(O_i)-\hat{\phi}_0(O_i)-\frac{R_i}{\pi_{\mathcal{R}}}\hat{\tau}^{\text{FB-Xonly}}\right\}^2.
\]
The asymptotic confidence interval is
\[
\left[\hat{\tau}^{\text{FB-Xonly}} - z_{1-\alpha/2}\sqrt{\hat{V}^{\text{FB-Xonly}}/n},\ \hat{\tau}^{\text{FB-Xonly}}+ z_{1-\alpha/2}\sqrt{\hat{V}^{\text{FB-Xonly}}/n}\right].
\]

\subsection*{A.3 Borrowing from auxiliary regions using inverse variance weighting (FB-IVW)}

% Recall that
% \begin{equation*}
% \hat{\tau}^{\text{FB-IVW}}=\hat{\theta}_1^{\text{FB-IVW}}-\hat{\theta}_0^{\text{FB-IVW}}.
% \end{equation*}
Let 
$$\psi_a(O)=\frac{R}{\pi_\mathcal{R}}\mu_{a}+\frac{\pi(X)}{\pi_\mathcal{R}}\frac{\mathbb{I}(A=a)}{e_a(X)}\{Y - \mu_{a}\},
$$
and 
$\hat{\psi}_a(O_i)$ denote the empirical version of $\psi_a(O)$ with estimated nuisance functions, that is, 
\begin{equation*}
\hat{\psi}_a(O_i)=\frac{R_i}{\pi_\mathcal{R}}\hat{Y}_i^{\text{FB}}+\frac{\hat{\pi}^{\text{FB}}(X_i)}{\pi_\mathcal{R}}\frac{\mathbb{I}(A_i=a)}{e_a(X_i)}
\{Y_i-\hat{Y}_i^{\text{FB}}\}.
\end{equation*}
where $$
\hat{Y}_i^{\text{FB}}
=
\begin{cases}
\dfrac{v_{\text{FB}}}{v_{\text{NB}}+v_{\text{FB}}}\hat{\mu}_a^{\text{NB}}(X_i,U_i)
+\dfrac{v_{\text{NB}}}{v_{\text{NB}}+v_{\text{FB}}}\hat{\mu}_a^{\text{FB}}(X_i),
& R_i=1, \\
\hat{\mu}_a^{\text{FB}}(X_i),
& R_i=0.
\end{cases}
$$
% The EIF of $\tau$ is
% \[
% \mathbb{IF}(\tau) 
% = \mathbb{IF}(\theta_1 - \theta_0) 
% = \mathbb{IF}(\theta_1) - \mathbb{IF}(\theta_0) 
% = \psi_1(O) - \psi_0(O) - \frac{R}{\pi_R}\hat{\tau}^{\text{FB-IVW}}.
% \]
The variance estimator for $\sqrt{n}\hat{\tau}^{\text{FB-IVW}}$ is
\[
\hat{V}^{\text{FB-IVW}} = \frac{1}{n}\sum_{i=1}^n \left\{\hat{\psi}_1(O_i)-\hat{\psi}_0(O_i)-\frac{R_i}{\pi_R}\hat{\tau}^{\text{FB-IVW}}\right\}^2.
\]
The asymptotic confidence interval is
\[
\left[\hat{\tau}^{\text{FB-IVW}} - z_{1-\alpha/2}\sqrt{\hat{V}^{\text{FB-IVW}}/n},\ \hat{\tau}^{\text{FB-IVW}}+ z_{1-\alpha/2}\sqrt{\hat{V}^{\text{FB-IVW}}/n}\right].
\]

\begin{remark}
Note that by rearranging terms, we can obtain the equivalent representation
\begin{align*}
\hat{\theta}^{\mathrm{FB\text{-}IVW}}_{a}
&=
\frac{1}{n_{\mathcal R}}
\sum_{i=1}^n
R_i\,\hat{\pi}^{\mathrm{FB}}(X_i)\,
\frac{\mathbb{I}(A_i = a)}{e_a(X_i)}\,Y_i
\\
&\quad+
\frac{1}{n_{\mathcal R}}
\sum_{i=1}^n
R_i
\left\{
1 -
\hat{\pi}^{\mathrm{FB}}(X_i)
\frac{\mathbb{I}(A_i = a)}{e_a(X_i)}
\right\}
\hat{Y}_i^{\mathrm{FB}}
\\
&\quad+
\frac{1}{n_{\mathcal R}}
\sum_{i=1}^n
(1-R_i)\,
\hat{\pi}^{\mathrm{FB}}(X_i)
\frac{\mathbb{I}(A_i = a)}{e_a(X_i)}
\left(Y_i-\hat{Y}_i^{\mathrm{FB}}\right).
\end{align*}

The first term corresponds to the estimator that would be obtained  based on the target region. Under Assumption 2, $\hat{Y}_i^{\mathrm{FB}}$ consistently estimates the conditional mean outcome, implying that the rest of the terms can be considered as an augmentation term has mean zero asymptotically. In this sense, $\hat{\theta}^{\mathrm{FB\text{-}IVW}}_{a}$ may be viewed as a target-only estimator plus a mean-zero augmentation term which is conceptually related to the efficiency-augmentation idea considered by \cite{yang2020combining}, where an estimator of zero is added to improve the efficiency of a target-only estimator.
\end{remark}

\section*{Appendix B: Algorithm for adaptive selection threshold of $\gamma_a$}

\begin{algorithm}
\caption{Adaptive Selection Threshold of $\gamma_a$}
\textbf{Input:} Threshold grid $\mathcal{G} = \{0, 0.1, \ldots, 1\}$, number of bootstrap samples $L = 100$.  
\begin{algorithmic}[1]
\For{$\gamma_a \in \mathcal{G}$}
    \State Compute $\hat{\theta}_a^{\text{CSB-IVW}}(\gamma_a)$ from the original sample.
    \For{$l = 1, \ldots, L$}
        \State Compute $\hat{\theta}_a^{\text{CSB-IVW} (l)}(\gamma_a)$ from the $l$-th bootstrap sample.
    \EndFor
\EndFor
\For{$\gamma_a \in \mathcal{G} \setminus \{1\}$}

    \State $\widehat{\mathbb{V}}\{\hat{\theta}_a^{\text{CSB-IVW}}(\gamma_a)
          - \hat{\theta}_a^{\text{NB-AllCov}}\}
    = (L - 1)^{-1} \sum_{l=1}^{L} \Bigg[
        \{\hat{\theta}_a^{\text{CSB-IVW}(l)}(\gamma_a)
        - \hat{\theta}_a^{\text{NB-AllCov}(l)}\}$
    \Statex \hspace{\algorithmicindent}$\quad - L^{-1} \sum_{l'=1}^{L}
        \{\hat{\theta}_a^{\text{CSB-IVW}(l')}(\gamma_a)
        - \hat{\theta}_a^{\text{NB-AllCov}(l')}\}
    \Bigg]^2$

   %\State $\widehat{\mathbb{V}}\{\hat{\theta}_a^{\text{CSB-IVW}}(\gamma_a)-\hat{\theta}_a^{\text{NB-AllCov}}\} = (L - 1)^{-1} \sum_{l=1}^{L} \left[\{\hat{\theta}_a^{\text{CSB-IVW}(l)}(\gamma_a)-\hat{\theta}_a^{\text{NB-AllCov}(l)}\}- L^{-1} \sum_{l'=1}^{L} \{\hat{\theta}_a^{\text{CSB-IVW}(l')}(\gamma_a)-\hat{\theta}_a^{\text{NB-AllCov}(l')}\} \right]^2$
    
    \State $\widehat{\mathbb{V}}\{\hat{\theta}_a^{\text{CSB-IVW}}(\gamma_a)\} = (L - 1)^{-1} \sum_{l=1}^{L} \left\{ \hat{\theta}_a^{\text{CSB-IVW}(l)}(\gamma_a) - L^{-1} \sum_{l'=1}^{L} \hat{\theta}_a^{\text{CSB-IVW}(l')}(\gamma_a) \right\}^2$
    
    \State $\widehat{\text{MSE}}_a(\gamma_a) = \{\hat{\theta}_a^{\text{CSB-IVW}}(\gamma_a)
          - \hat{\theta}_a^{\text{NB-AllCov}}\}^2 - \widehat{\mathbb{V}}\{\hat{\theta}_a^{\text{CSB-IVW}}(\gamma_a)
          - \hat{\theta}_a^{\text{NB-AllCov}}\} + \widehat{\mathbb{V}}\{\hat{\theta}_a^{\text{CSB-IVW}}(\gamma_a)\}$
\EndFor
\State $\widehat{\text{MSE}}_a(1) = (L - 1)^{-1} \sum_{l=1}^{L} \left\{ \hat{\theta}_a^{\text{NB-AllCov}(l)} - L^{-1} \sum_{l'=1}^{L} \hat{\theta}_a^{\text{NB-AllCov}(l')} \right\}^2$
\State $\gamma_a^{*} = \arg\min_{\gamma_a \in \mathcal{G}} \widehat{\text{MSE}}_a(\gamma_a)$
\State \textbf{Output:} $\gamma_a^{*}$
\end{algorithmic}

\label{alg1}

\end{algorithm}

\newpage
\section*{Appendix C: Proof of Theorem 1} 

Under $H_0$, the imputed potential outcomes are the same as the true potential
outcomes. Thus, the distribution of $T^* \equiv T(\boldsymbol{A}^*)$ is the same as that of
$T \equiv T(\boldsymbol{A})$ given the subassignment vector corresponding to auxiliary-region observations $\boldsymbol{A}_{\mathcal{E}}$, which are held fixed when constructing the randomization distribution over the treatment assignments $\boldsymbol{A}$ in the target region. With simplified notations, we have
\[
\mathbb{P}_{\boldsymbol{A}}\bigl(p^{\mathrm{FRT}}\le \alpha\bigr |{\boldsymbol{A}_{\mathcal{E}}})
  = \mathbb{P}_{\boldsymbol{A}}\bigl\{\mathbb{P}_{{\boldsymbol{A}}^*}(T^* \ge T) \le \alpha\bigr |{\boldsymbol{A}_{\mathcal{E}}}\}.
\]

In a finite sample, $\boldsymbol{A}$ can take only a finite set of values, which implies that
$T$ must also take on a finite set of values. Suppose these values are
\[
  T_1 > \cdots > T_m > \cdots > T_M,
\]
and
\[
  \mathbb{P}_{\boldsymbol{A}}(T = T_m |{\boldsymbol{A}_{\mathcal{E}}}) = \mathbb{P}_{{\boldsymbol{A}}^*}(T^* = T_m |{\boldsymbol{A}_{\mathcal{E}}}) = \alpha_m,
  \qquad m = 1,\ldots,M.
\]
For $T \in \{T_1,\ldots,T_M\}$, we have
$\alpha_1 \le \mathbb{P}_{{\boldsymbol{A}}^*}(T^* \ge T) \le \sum_{m=1}^M \alpha_m = 1$.
If $0 < \alpha < \alpha_1$, we have
\[
  \mathbb{P}_{\boldsymbol{A}}\bigl(p^{\mathrm{FRT}} \le \alpha\bigr|)
  = \mathbb{P}_{\boldsymbol{A}}\bigl\{\mathbb{P}_{{\boldsymbol{A}}^*}(T^* \ge T) \le \alpha\bigr |{\boldsymbol{A}_{\mathcal{E}}} \}
  = 0 \le \alpha.
\]

If $\alpha_1 \le \alpha < 1$, $\exists\,\tilde M \in \{1,\ldots,M-1\}$ such that
$\sum_{m=1}^{\tilde M} \alpha_m \le \alpha$ and
$\sum_{m=1}^{\tilde M+1} \alpha_m > \alpha$. Then, we have
\[
  \mathbb{P}_{\boldsymbol{A}}\bigl(p^{\mathrm{FRT}} \le \alpha\bigr |{\boldsymbol{A}_{\mathcal{E}}})
  = \mathbb{P}_{\boldsymbol{A}}\bigl\{\mathbb{P}_{A^*}(T^* \ge T) \le \alpha\bigr |{\boldsymbol{A}_{\mathcal{E}}}\}
  = \mathbb{P}_{\boldsymbol{A}}\{T \in \{T_1,\ldots,T_{\tilde M}\}|{\boldsymbol{A}_{\mathcal{E}}}\}
  = \sum_{m=1}^{\tilde M} \alpha_m \le \alpha.
\]

If $T(\boldsymbol{A})$ takes distinct values for different $\boldsymbol{A} \in \mathcal{A}$,
$p^{\mathrm{FRT}}$ is uniformly distributed:
\[
  \mathbb{P}_{\boldsymbol{A}}\left(p^{\mathrm{FRT}} = \frac{a}{|\mathcal{A}|}|{\boldsymbol{A}_{\mathcal{E}}}\right)
  = \frac{1}{|\mathcal{A}|}, \qquad a = 1,\ldots,|\mathcal{A}|.
\]
Thus, we have
\[
  \mathbb{P}_{\boldsymbol{A}}\bigl(p^{\mathrm{FRT}} \le \alpha\bigr|{\boldsymbol{A}_{\mathcal{E}}})
  = \frac{\lfloor \alpha|\mathcal{A}|\rfloor}{|\mathcal{A}|}
  > \frac{\alpha|\mathcal{A}| - 1}{|\mathcal{A}|}
  = \alpha - \frac{1}{|\mathcal{A}|}.
\]

We can remove the conditioning on ${\boldsymbol{A}_{\mathcal{E}}}$ by averaging over its randomization distribution. By the law of total probability, we have
\[
\mathbb{P}_{\boldsymbol{A}}\!\left(p^{\mathrm{FRT}} \le \alpha\right)
  = \sum_{{\boldsymbol{A}_{\mathcal{E}}}} \mathbb{P}_{\boldsymbol{A}}\!\left(p^{\mathrm{FRT}} \le \alpha, {\boldsymbol{A}_{\mathcal{E}}} \right)
  = \sum_{{\boldsymbol{A}_{\mathcal{E}}}} \mathbb{P}_{\boldsymbol{A}}\!\left(p^{\mathrm{FRT}} \le \alpha \mid {\boldsymbol{A}_{\mathcal{E}}} \right)\mathbb{P}({\boldsymbol{A}_{\mathcal{E}}}),
\]
where the sum is taken over all possible randomized
auxiliary data.  

From part (i), under the sharp null hypothesis $H_0$ we have, for every ${\boldsymbol{A}_{\mathcal{E}}}$,
\[
\mathbb{P}_{\boldsymbol{A}}\!\left(p^{\mathrm{FRT}} \le \alpha \mid {\boldsymbol{A}_{\mathcal{E}}} \right) \le \alpha.
\]
Substituting this bound into the previous display yields
\[
\mathbb{P}_{\boldsymbol{A}}\!\left(p^{\mathrm{FRT}} \le \alpha\right)
  = \sum_{{\boldsymbol{A}_{\mathcal{E}}}} \mathbb{P}_{\boldsymbol{A}}\!\left(p^{\mathrm{FRT}} \le \alpha \mid {\boldsymbol{A}_{\mathcal{E}}} \right)\mathbb{P}({\boldsymbol{A}_{\mathcal{E}}})
  \le \sum_{{\boldsymbol{A}_{\mathcal{E}}}} \alpha\,\mathbb{P}({\boldsymbol{A}_{\mathcal{E}}})
  = \alpha \sum_{{\boldsymbol{A}_{\mathcal{E}}}} \mathbb{P}({\boldsymbol{A}_{\mathcal{E}}})
  = \alpha.
\]

% \noindent\textit{Remark C.1.}
% If $T$ is a continuous random variable, suppose its distribution function is
% $F(t)=P(T\le t)$, then the proof could be simplified as
% \begin{align*}
% \mathbb{P}_{\boldsymbol{A}}\!\left\{\mathbb{P}_{{\boldsymbol{A}}^*}(T^*\ge T)\le \alpha\right |{\boldsymbol{A}_{\mathcal{E}}}\}
%   &= P\{1-F(T)\le \alpha |{\boldsymbol{A}_{\mathcal{E}}}\} \\
%   &= P\{T \ge F^{-1}(1-\alpha)|{\boldsymbol{A}_{\mathcal{E}}}\} \\
%   &= 1 - F\!\left(F^{-1}(1-\alpha)\right) \\
%   &= \alpha.
% \end{align*}

% \noindent
% However, $T$ is discrete with finite values, and we provide a rigorous proof in the finite-sample setting.

\section*{Appendix D: Additional simulation details}

\subsection*{D.1: Simulation data-generating parameters}

Table \ref{parameter} presents \rev{a summary of the data-generating process}.

\begin{table}[htbp]
\centering

\caption{Simulation data-generating parameters.}
\label{tab:sim-parameters}
\resizebox{\textwidth}{!}{%
\begin{tabular}{lll}
\hline
\textbf{Parameter} & \textbf{Description} & \textbf{Values / Specification} \\
\hline
$\{n_{\mathcal{R}},n_{\mathcal{E},}\}$ & Sample sizes in target and auxiliary regions & \{600, 1000\} \\[4pt]
$X,U = (X_1, X_2,U)$ & Covariates & (i) $  \mathcal{MVN}( \big(\begin{smallmatrix} 0 \\ 0 \\2 \end{smallmatrix} \big),\big(\begin{smallmatrix} 1 & 0 & 0.5 \\ 0 & 1 & 0.5 \\ 0.5 & 0.5 & 1 \end{smallmatrix} \big)$ or (ii) $  \mathcal{MVN}( \big(\begin{smallmatrix} 0 \\ 0 \\2 \end{smallmatrix} \big),\big(\begin{smallmatrix} 1 & 0 & 0 \\ 0 & 1 & 0 \\ 0 & 0 & 1 \end{smallmatrix} \big)$ \\[4pt]
$\pi_1(X)$ & Sampling probability into the target region &
$ \pi_1(X) = \{1 + \exp(X^\top \eta_1 + 0.6)\}^{-1} $ with  $\eta_1 = (0.5, 0.3)$;\\[4pt]
$\pi_0(X)$ & Sampling probability into the auxiliary region &
$ \pi_0(X) = \{1 + \exp(X^\top \rev{\eta_0} - 0.4)\}^{-1} $ with  $\rev{\eta_0} = (-0.5, -0.2)$;\\[4pt]
$R$ & Sampling indicator & $\text{Bernoulli}(\pi(X))$ with $\pi(X)=\frac{\pi_1(X)}{\pi_1(X)+\pi_0(X)}$\\[4pt]
$A$ & Treatment assignment & $\text{Bernoulli}(0.5)$ within each region \\[4pt]
$\beta_a$ & Regression coefficients in outcome model &
$\beta_0 = (0, 2, 2)$; $\beta_1 = (3, 3, 3)$ \\[4pt]
$\alpha_a$ & Coefficient for unobserved $U$ &
$\alpha_0 = \{0.1, 0.5, 1, 1.5\}$; $\alpha_1 = 2\alpha_0$ \\[4pt]
$\varepsilon$ & Error variance & $\varepsilon \in \{0.1, 0.5, 1, 1.5\}$ \\[4pt]
$b_0, b_1$ & Hidden bias in auxiliary region & $b_0 = 6$; $b_1 = 10$ \\[4pt]
$\rho$ & Proportion of biased auxiliary samples & $50\%$ \\[4pt]

\hline
\end{tabular}
}\label{parameter}
\end{table}

\subsection*{D.2: Simulation results for MSE\% vs log(Precision) across different signals of $X$ with independent $X$ and $U$.}

\begin{figure}[htbp]
    \centering   \includegraphics[width=\textwidth]
    {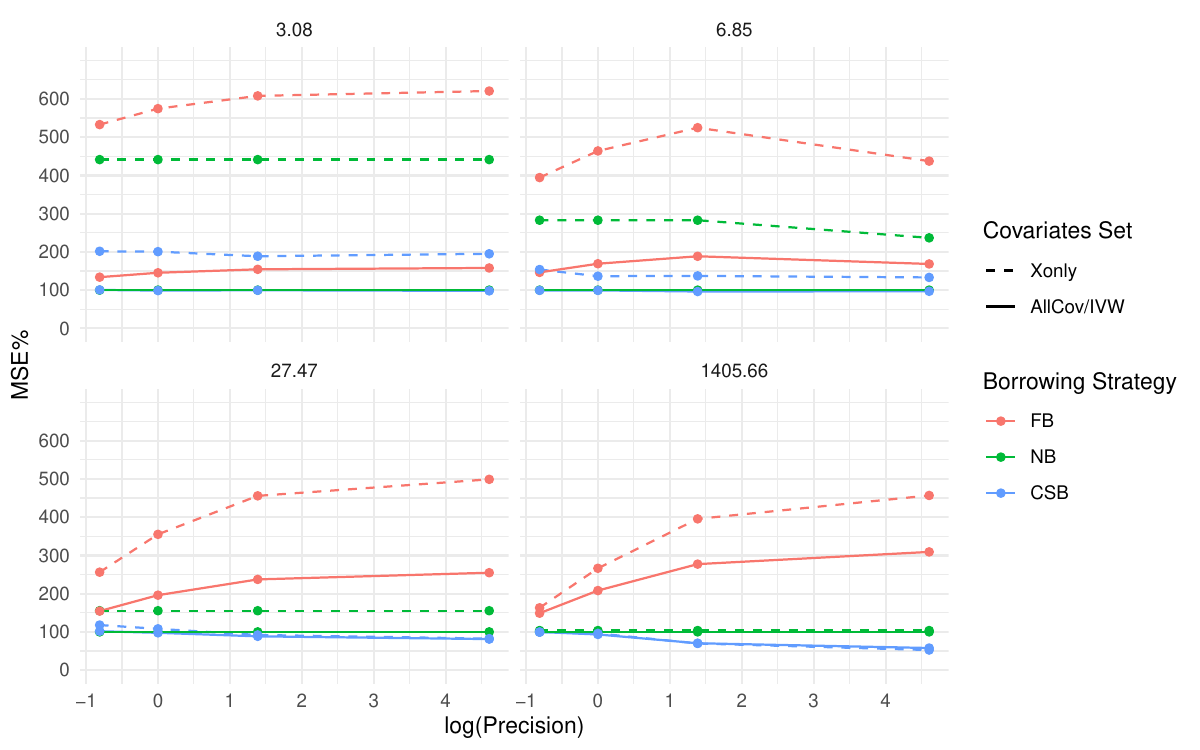}
    \caption{MSE\% vs log(Precision) across different signals of $X$ with independent $X$ and $U$.}
    \label{figure1}
\end{figure}

Figures \ref{figure1} present the performance of the six estimators under varying levels of
precision and signal strength of the observed covariates $X$ when $X$ and $U$ are independent.

\section*{Appendix E: More details about the real data}
\subsection*{E.1: Baseline patient characteristics by treatment group across regions}
Detailed baseline characteristics after matching across arms are summarized in Table \ref{table2}.

\begin{table}[htbp]
    \caption{Baseline patient characteristics by treatment group across regions. ($^\dagger$Median (Q1, Q3); n (\%)).}
    \centering
    \resizebox{\textwidth}{!}{%
    \begin{tabular}{lccccc}
        \Hline
        \textbf{Characteristic} &
        \shortstack{\textbf{Overall}\\(N = 345$^{\dagger}$)} &
        \shortstack{\textbf{Auxiliary Regions} \\ \textbf{ GTx-024 3mg}\\(N = 121$^{\dagger}$)} &
        \shortstack{\textbf{Auxiliary Regions} \\ \textbf{Placebo}\\(N = 155$^{\dagger}$)} &
        \shortstack{\textbf{North America}\\ \textbf{ GTx-024 3mg}\\(N = 39$^{\dagger}$)} &
        \shortstack{\textbf{North America}\\ \textbf{ Placebo}\\(N = 30$^{\dagger}$)} \\
        \hline
        \textbf{Age} & 65 (60, 71) & 64 (60, 70) & 65 (60, 70) & 64 (59, 72) & 67 (57, 72) \\ 
        \textbf{Sex} & & & & & \\
        \quad F & 147 (43\%) & 50 (41\%) & 67 (43\%) & 17 (44\%) & 13 (43\%) \\
        \quad M & 198 (57\%) & 71 (59\%) & 88 (57\%) & 22 (56\%) & 17 (57\%) \\
        \textbf{ECOG baseline} & & & & & \\
        \quad 0 – normal activity; asymptomatic & 135 (39\%) & 43 (36\%) & 64 (41\%) & 19 (49\%) & 9 (30\%) \\
        \quad 1 – symptomatic; fully ambulatory & 210 (61\%) & 78 (64\%) & 91 (59\%) & 20 (51\%) & 21 (70\%) \\
        \textbf{Chemotherapy} & & & & & \\
        \quad Non-taxane & 165 (48\%) & 60 (50\%) & 72 (46\%) & 17 (44\%) & 16 (53\%) \\
        \quad Taxane & 180 (52\%) & 61 (50\%) & 83 (54\%) & 22 (56\%) & 14 (47\%) \\
        \textbf{Cancer stage} & & & & & \\
        \quad Stage III & 106 (31\%) & 42 (35\%) & 44 (28\%) & 11 (28\%) & 9 (30\%) \\
        \quad Stage IV & 239 (69\%) & 79 (65\%) & 111 (72\%) & 28 (72\%) & 21 (70\%) \\
        \textbf{Histology} & & & & & \\
        \quad non-squamous cell carcinoma & 213 (62\%) & 71 (59\%) & 96 (62\%) & 22 (56\%) & 24 (80\%) \\
        \quad squamous cell carcinoma & 132 (38\%) & 50 (41\%) & 59 (38\%) & 17 (44\%) & 6 (20\%) \\
        \textbf{LBM baseline} & 44 (37, 52) & 45 (38, 51) & 44 (38, 52) & 43 (36, 53) & 40 (37, 52) \\
        \textbf{Weight loss flag} & & & & & \\
        \quad 0 & 32 (46\%) & 0 (NA\%) & 0 (NA\%) & 17 (44\%) & 15 (50\%) \\
        \quad 1 & 37 (54\%) & 0 (NA\%) & 0 (NA\%) & 22 (56\%) & 15 (50\%) \\
        \quad Unknown & 276 & 121 & 155 & 0 & 0 \\
        \Hline
    \end{tabular}
    } % end resizebox
    \label{table2}
\end{table}

\subsection*{E.2: Covariate balance}
\begin{figure}[htbp]
    \centering   \includegraphics[width=\textwidth]
    {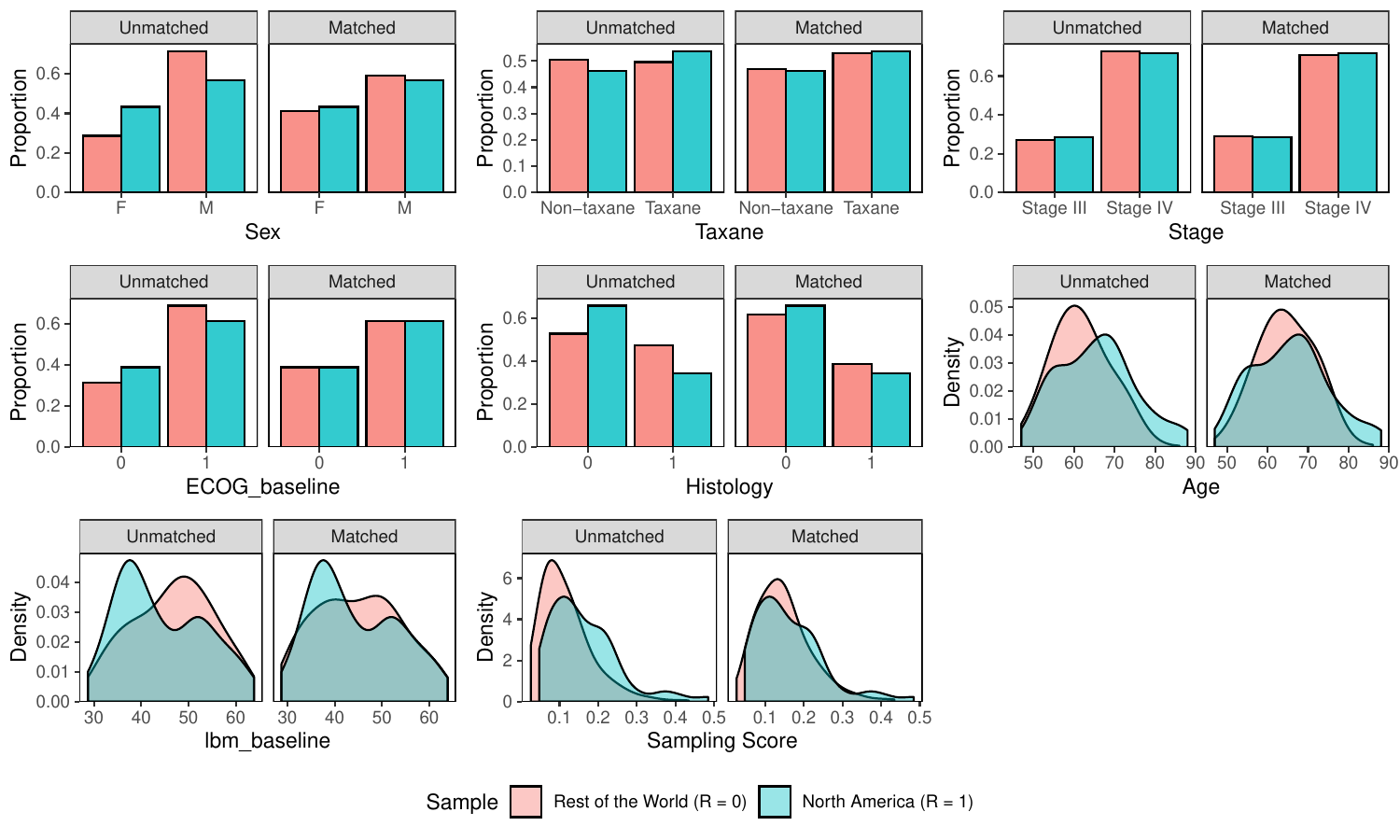}
    \caption{Distributional balance (unmatched and matched) between North America ($R = 1$) and auxiliary regions ($R = 0$) for baseline covariates and the estimated sampling score $\hat{\mathbb{P}}(R=1|X)$.}
    \label{figure5}
\end{figure}

Figure \ref{figure5} showed that the distributional balance for covariates and sampling score improved substantially after matching. However, imbalances persisted in variables such as age and baseline LBM, with North American patients generally older and with lower baseline LBM. These residual imbalances could not be resolved through matching without discarding North American samples, which motivates the use of the doubly robust estimators proposed in Sections 3 and 4.

\section*{F: Extension to multiple auxiliary regions with distinct \rev{covariate} sets}
\label{app:multi-region}

\subsection*{F.1 Data structure}

Let the region indicator satisfy
$R \in \{1\} \cup \mathcal{B}_0$,
where $R=1$ indexes the target region and
$\mathcal{B}_0=\{r_1,\ldots,r_S\}$ indexes the $S$ auxiliary regions.
The observed data remain
\[
\{R_i, X_i, U_i, A_i, Y_i\}_{i=1}^n,
\]
with $U_i$ observed only when $R_i=1$. For each auxiliary region $r\in\mathcal{B}_0$, let
$X^{(r)} \subseteq X$ denote the subset of baseline covariates that are jointly observed and well-defined in both
the target region ($R=1$) and auxiliary region $r$. These shared covariate sets are allowed
to differ across auxiliary regions. The covariate vector $X$ in the main text can be viewed
as the collection of all baseline covariates observed in at least one auxiliary region, while
borrowing from region $r$ conditions only on $X^{(r)}$. The estimand of interest remains the target-region region-specific average treatment effect $\tau = \mathbb{E}\{Y(1)-Y(0)\mid R=1\}$.

\subsection*{F.2 Region-specific identification and FB-IVW estimation}

We generalize Assumption~2 to a region-specific form. 

\setcounter{assumption}{2}
\begin{assumption}[Region-specific conditional mean exchangeability]
\label{ass:cme1}
For each auxiliary region $r\in\mathcal{B}_0$ and treatment arm $a\in\{0,1\}$,
\[
\mathbb{E}\{Y(a)\mid R=r, X^{(r)}\}
=
\mathbb{E}\{Y(a)\mid R=1, X^{(r)}\}.
\]
\end{assumption}

This assumption requires conditional mean exchangeability only between the target region
and a given auxiliary region, conditional on their shared covariates $X^{(r)}$, and does not
require a single covariate set common to all regions.

With Assumption \ref{ass:cme1}, using data from the target region and auxiliary region $r$, we fit
$\widehat{\mu}_{a}^\mathrm{NB}(X,U)$ using only target-region data and
$\widehat{\mu}_{a}^{,\mathrm{FB}(r)}(X^{(r)})$ using pooled data from regions $R=1$ and
$R=r$.

The IVW prediction is then defined as
$$
\hat{Y}_i^{\text{FB}(r)}
=
\begin{cases}
\dfrac{v_{\text{FB}}^{(r)}}{v_{\text{NB}}+v_{\text{FB}}^{(r)}}\hat{\mu}_a^{\text{NB}}(X_i,U_i)
+\dfrac{v_{\text{NB}}}{v_{\text{NB}}+v_{\text{FB}}^{(r)}}\hat{\mu}_a^{\text{FB}^{(r)}}(X_i^{(r)}),
& R_i=1, \\
\hat{\mu}_a^{\text{FB}^{(r)}}(X_i^{(r)}),
& R_i=r,
\end{cases}
$$
where
$$
v_{\text{NB}}
=\frac{\sum_{i=1}^n \mathbb{I}(A_i=a,R_i=1)\{Y_i-\hat{\mu}_a^{\text{NB}}(X_i,U_i)\}^2}
{\sum_{i=1}^n \mathbb{I}(A_i=a,R_i=1)},
$$
$$
v_{\text{FB}}^{(r)}
=\frac{\sum_{i=1}^n \mathbb{I}(A_i=a,R_i\in\{1,r\})\{Y_i-\hat{\mu}_a^{\text{FB}(r)}(X_i^{(r)})\}^2}
{\sum_{i=1}^n \mathbb{I}(A_i=a, R_i\in\{1,r\})}.
$$
Analogous to the estimator in Section 3, the corresponding region-specific full-borrowing IVW estimator is
$$
\hat{\theta}_a^{\mathrm{FB\text{-}IVW}(r)}
=
\frac{1}{n_R}
\sum_{i=1}^n
\left[
\mathbb{I}(R_i=1)\hat{Y}_i^{\mathrm{FB}(r)}
+
\hat{\pi}^{FB(r)}(X_i^{(r)})
\frac{\mathbb{I}(A_i=a,\ R_i\in\{1,r\})}
{e_a(X_i^{(r)})}
\{Y_i-\hat{Y}_i^{\mathrm{FB}(r)}\}
\right].
$$

When all auxiliary regions are pooled and share the same covariate set,
$\hat{\theta}_a^{\mathrm{FB\text{-}IVW}(r)}$ reduces to the estimator
$\hat{\theta}_a^{\mathrm{FB\text{-}IVW}}$ defined in Section 3. With multiple auxiliary regions available, each region $r\in\mathcal{B}_0$ yields a
region-specific estimator $\hat{\theta}_a^{\mathrm{FB\text{-}IVW}(r)}$. Then, the corresponding RSATE using data from region $r$ is $\hat{\tau}^{\mathrm{FB\text{-}IVW}(r)}=\hat{\theta}_1^{\mathrm{FB\text{-}IVW}(r)}-\hat{\theta}_0^{\mathrm{FB\text{-}IVW}(r)}$.

We can further combine these estimators using inverse-variance weighting \citep{gao2025improving}. Let $\hat{{\tau}}^{\mathrm{FB\text{-}IVW(r_1:r_S)}}
=
\bigl(
\hat{\tau}^{\mathrm{FB\text{-}IVW}(r_1)},
\ldots,
\hat{\tau}^{\mathrm{FB\text{-}IVW}(r_S)}\bigr)^\top$ denote the concatenated region-specific FB-IVW estimators corresponding to the
auxiliary regions $B_0=\{r_1,\ldots,r_S\}$, we have
$n^{1/2}
\left(
\hat{{\tau}}^{\mathrm{FB\text{-}IVW(r_1:r_S)}}
-
\tau
\right) \xrightarrow\
N({0}, \Sigma)$ where $\Sigma=\mathbb{E}[\psi^{(r_1:r_S)}(O)\{\psi^{(r_1:r_S)}(O)\}^\top]$ and  $\psi^{(r_1:r_S)}$ denote the influence functions of $\hat{{\tau}}^{\mathrm{FB\text{-}IVW(r_1:r_S)}}$. Thus, the final integrative estimator that combines information across all auxiliary regions
is
$\hat{\tau}^{\mathrm{FB\text{-}IVW}(final)}
=
\hat{d}^\top
\hat{{\tau}}^{\mathrm{FB\text{-}IVW(r_1:r_S)}} \xrightarrow\
N\!\left(
0,\,
\hat{d}^\top \Sigma \hat{d}
\right),
$
where the optimal integrative weight $\hat{d}$ to combine $\hat{{\tau}}^{\mathrm{FB\text{-}IVW(r_1:r_S)}}$ can be obtained by minimizing the post-integration \rev{variance}: min ${d}^\top \Sigma{d}$, subject to $d^\top 1_S=1$. By Lagrange multiplier method, we can show that $\hat{d}
=
\left\{
{1}_S^\top \Sigma^{-1}{1}_S
\right\}^{-1}
\Sigma^{-1}{1}_S$ is the optimal combining weights.

\subsection*{F.3 Conformal selective borrowing by region}

When evidence suggests that Assumption \ref{ass:cme1} is violated, we can apply CSB 
for each auxiliary region following Section~4.1. For an auxiliary observation $j$ with $R_j=r\in\mathcal{B}_0$ and $A_j=a$, the
conformal $p$-value $p_j$ is obtained by comparing $(X_j^{(r)},Y_j)$ to the target-region
observations $\{(X_i,Y_i): R_i=1, A_i=a\}$. For each treatment arm $a$ and auxiliary region $r$, define the selected auxiliary set
\[
\mathcal{E}_a^{(r)}(\gamma_a^{(r)})
=
\{j: R_j=r,\ A_j=a,\ p_j\ge\gamma_a^{(r)}\}.
\]
The CSB--IVW estimator is then constructed using the target-region
data together with the pooled selected auxiliary observations
$\bigcup_{r\in\mathcal{B}_0} \mathcal{E}_a^{(r)}(\gamma_a^{(r)})$,
with outcome regressions, sampling scores, and IVW predictions defined region by region on
$X^{(r)}$. Thresholds $\gamma_a^{(r)}$ may be selected separately across auxiliary
regions using the same MSE-guided procedure as in the main text.

\label{lastpage}

\bibliographystyle{biom}
\bibliography{_ref}